\documentclass[fleqn,usenatbib]{mnras}

\usepackage{newtxtext,newtxmath}

\usepackage[T1]{fontenc}
\usepackage{longtable}

\DeclareRobustCommand{\VAN}[3]{#2}
\let\VANthebibliography\thebibliography
\def\thebibliography{\DeclareRobustCommand{\VAN}[3]{##3}\VANthebibliography}

\usepackage{graphicx}	
\usepackage{amsmath}	
\usepackage{longtable}
\usepackage{orcidlink}

\def\arcmin{\hbox{$^\prime$}}
\def\arcsec{\hbox{$^{\prime\prime}$}}

\title[The nature of EP~241021a]{Unveiling the nature of the Einstein Probe transient EP~241021a}

\author[Quirola-V\'asquez et al.]{
J. Quirola-V\'asquez,\orcidlink{0000-0001-8602-4641} $^{1}$\thanks{E-mail: jonathan.quirolavasquez@ru.nl}
P.~G.~ Jonker,\orcidlink{0000-0001-5679-0695} $^{1}$
A. J. Levan,\orcidlink{0000-0001-7821-9369} $^{1,2}$
D. B. Malesani,\orcidlink{0000-0002-7517-326X} $^{3,4}$
F. E. Bauer,\orcidlink{0000-0002-8686-8737} $^{5}$
N. Sarin,\orcidlink{0000-0003-2700-1030} $^{6}$
\newauthor{G. P. Lamb},\orcidlink{0000-0001-5169-4143} $^{7}$
A. Martin-Carrillo,\orcidlink{0000-0001-5108-0627} $^{8}$
J. S\'anchez-Sierras,\orcidlink{0000-0003-2276-4231} $^{1}$
M. Fraser,\orcidlink{0000-0003-2191-1674} $^{8}$
L. Izzo,\orcidlink{0000-0001-9695-8472} $^{9,10}$
M. E. Ravasio,\orcidlink{0000-0003-3193-4714} $^{1}$
\newauthor{D. Mata S\'anchez},\orcidlink{0000-0003-0245-9424} $^{11,12}$
M. A. P. Torres,\orcidlink{0000-0002-5297-2683} $^{11,12}$
J. N. D. van Dalen,\orcidlink{0009-0007-6927-7496} $^{1}$
A. P. C. van Hoof,\orcidlink{0009-0005-5404-2745} $^{1}$
J. A. Chac\'on,\orcidlink{0009-0000-6374-3221} $^{13,14}$
\newauthor{S. Littlefair},\orcidlink{0000-0001-7221-855X} $^{15}$
V. S. Dhillon,\orcidlink{0000-0003-4236-9642} $^{15,11}$
L. Cotter,\orcidlink{0000-0002-7910-6646} $^{8}$
G. Corcoran,\orcidlink{0009-0009-1573-8300} $^{8}$
R. A. J. Eyles-Ferris,\orcidlink{0000-0002-8775-2365} $^{16}$
P. T. O’Brien,\orcidlink{0000-0002-5128-1899} $^{16}$
\newauthor{D. Stern},\orcidlink{0000-0003-2686-9241} $^{17}$
F. Harrison,\orcidlink{0000-0002-4226-8959} $^{18}$
V. D'Elia,\orcidlink{0000-0003-3703-4418} $^{19}$
D. H. Hartmann,\orcidlink{0000-0002-8028-0991} $^{20}$
\\
$^{1}$Department of Astrophysics/IMAPP, Radboud University Nijmegen, P.O.~Box 9010, Nijmegen, 6500~GL, The Netherlands\\
$^{2}$University of Warwick, Department of Physics, Coventry, CV4 7AL, United Kingdom\\
$^{3}$The Cosmic Dawn Centre (DAWN), Denmark\\
$^{4}$Niels Bohr Institute, University of Copenhagen, Jagtvej 155, DK-2200, Copenhagen N, Denmark\\
$^{5}$Instituto de Alta Investigaci{\'{o}}n, Universidad de Tarapac{\'{a}}, Casilla 7D, Arica, Chile\\
$^{6}$The Oskar Klein Centre, Department of Physics, Stockholm University, AlbaNova, Stockholm, SE-106 91, Stockholm, Sweden\\
$^{7}$Astrophysics Research Institute, Liverpool John Moores University, IC 2, Liverpool Science Park, 146 Brownlow Hill, Liverpool L3 5RF, UK\\
$^{8}$School of Physics and Centre for Space Research, University College Dublin, Belfield, Dublin 4, Ireland\\
$^{9}$Osservatorio Astronomico di Capodimonte, INAF, Salita Moiariello 16, Napoli, 80131, Italy\\
$^{10}$Niels Bohr Institute, University of Copenhagen, DARK, Jagtvej 128, Copenhagen, 2200, Denmark\\
$^{11}$Instituto de Astrof\'isica de Canarias, E-38205, La Laguna, Tenerife, Spain\\
$^{12}$Universidad de La Laguna, Departamento de Astro\'isica,  E-38206, La Laguna, Tenerife, Spain\\
$^{13}$Instituto de Astrof{\'{\i}}sica, Facultad de F{\'{i}}sica, Pontificia Universidad Cat{\'{o}}lica de Chile, Campus San Joaquín, Av. Vicuña Mackenna 4860, \\
Macul Santiago, Chile, 7820436\\
$^{14}$Millennium Institute of Astrophysics, Nuncio Monse{\~{n}}or S{\'{o}}tero Sanz 100, Of 104, Providencia, Santiago, Chile\\
$^{15}$Astrophysics Research Cluster, School of Mathematical and Physical Sciences, University of Sheffield, Sheffield, S3 7RH, UK.\\
$^{16}$School of Physics and Astronomy, University of Leicester, University Road, Leicester, LE1 7RH, UK\\
$^{17}$Jet Propulsion Laboratory, California Institute of Technology, 4800 Oak Grove Drive, Mail Stop 264-789, Pasadena, CA 91109, USA.\\
$^{18}$Department of Astronomy, California Institute of Technology, 1200 East California Blvd, Pasadena, CA 91125, USA.\\
$^{19}$Space Science Data Center (SSDC) - Agenzia Spaziale Italiana (ASI), Via del Politecnico snc, I-00133 Roma, Italy \\
$^{20}$Clemson University, Department of Physics \& Astronomy, Clemson, SC 29634, USA
}

\date{Accepted XXX. Received YYY; in original form ZZZ}

\pubyear{\the\year{}}

\begin{document}
\label{firstpage}
\pagerange{\pageref{firstpage}--\pageref{lastpage}}
\maketitle

\begin{abstract}
We present a multi-wavelength analysis of the fast X-ray transient EP~241021a, discovered by the Wide-field X-ray Telescope aboard the \emph{Einstein Probe} satellite on 2024 October 21. The event was not detected in gamma-rays. Follow-up observations from $\sim$1.5 to 100 days post-trigger were obtained across X-ray, UV, optical, near-infrared, and radio bands with ground- and space-based facilities. The redshift is constrained to $z = 0.7485$ from prominent optical spectral features.
The optical light curve shows complex evolution: an initial $\sim t^{-0.7}$ decay, followed by a rapid re-brightening peaking at day 7.7 with $\sim t^{-1.7}$ decay, and a third phase peaking near day 19 with $\sim t^{-1.3}$ decay. The spectral energy distribution (SED) and its temporal evolution are consistent with a mix of non-thermal and thermal components. Early optical-to-X-ray spectral indices agree with optically thin synchrotron emission, while steepening of the optical SED after $\sim$20 days indicates either a shift in emission mechanism or the emergence of an additional component.
Although broad-lined absorption features are absent, comparisons with type Ic-BL supernovae suggest a SN contribution at late times, suggesting a collapsar origin for EP~241021a. 
The likely SN in EP~241021a appears to require an additional energy source beyond $^{56}$Ni decay. These results support the view that some fast X-ray transients detected by the \emph{Einstein Probe} arise from massive stellar explosions.

\end{abstract}




\section{Introduction}


Extragalactic fast X-ray transients (FXTs) are defined as bursts of soft X-ray photons (in the ${\sim}$0.3--10~keV band) with durations from seconds to hours \citep{Heise2010}.
Over the decades, our knowledge of FXTs has increased substantially via the identification and characterization of sources detected by the new generation of X-ray satellites such as the Neil Gehrels \emph{Swift} Observatory \citep[\emph{Swift}; e.g.,][]{Soderberg2008,Evans2023}, the \emph{Chandra} X-ray Observatory \citep[e.g.,][]{Jonker2013,Glennie2015,Irwin2016,Bauer2017,Xue2019,Lin2022,Quirola2022,Quirola2023}, and the \emph{X-ray Multi-mirror Mission - Newton telescope} \citep[\emph{XMM-Newton}; e.g.,][]{Novara2020,Alp2020,DeLuca2021}. 
The high spatial resolution of those X-ray telescopes has mainly permitted ruling out any association with stellar flares, i.e., strengthening their extragalactic nature, as well as enabling the localization of the host galaxies of the transients \citep[e.g.,][]{Lin2022,Eappachen2022,Eappachen2023a,Eappachen2024,Quirola2024b}¡.

Several different progenitor models have been suggested to explain the X-ray light curve and spectra of extragalactic FXTs, as well as the multi-wavelength emission of contemporaneous counterparts: 
$i)$ the shock breakout (SBO) associated with certain types of core-collapse supernovae (CC-SNe) could produce an FXT, when the shock wave generated by a CC-SN reaches the surface of the progenitor star \citep[e.g., SN2008D/XRF 080109;][]{Soderberg2008,Waxman2017,Alp2020,Sun2022,Scully2023}; 
$ii)$ some models involving binary neutron star (BNS) mergers lead to the formation of a millisecond spin period highly magnetized neutron star -- a magnetar -- that powers a nearly isotropic X-ray signal detectable as an FXT \citep[e.g., XRT~141001, XRT~150322, and recently EP~250207b;][]{Dai2006,Metzger2008,Yu2013,Zhang2013,Sun2017,Sun2019,Quirola2024,Jonker2025}; 
$iii)$ recently, some FXTs have shown a connection with luminous fast blue optical transients (LFBOTs), and/or broad-lined Type Ic SNe \citep[e.g., EP~240414a and EP~250108a;][]{van_Dalen2024,Srivastav2025,Rastinejad2025,Eyles_Ferris2025,Srinivasaragavan2025,Li2025b,Sun2024};
$iv)$ a subset of FXTs may originate from the tidal disruption of a white dwarf (WD) passing close to an intermediate-mass black hole \citep[IMBH, $\sim10^3$–$10^5$ M$_\odot$; e.g., XRT~000519 and  EP~240408a;][]{Jonker2013,MacLeod2016,Maguire2020,OConnor2025}; and
$v)$ high-redshift and low-luminous gamma ray bursts (GRB) would naturally be redshifted to the X-ray range by cosmic expansion \citep[e.g., EP~240315a and EP~240801a;][]{Levan2024,Gillanders2024,Li2025b,Jiang2025}.
Nevertheless, until 2024, in the vast majority of cases, the X-ray transients themselves have only been identified long after the X-ray trigger, rendering contemporaneous follow-up observations impossible. Hence, determining their energetics and distance scale, and, by extension, their physical origin, remained a challenge.

The \emph{Einstein Probe} \citep[EP;][]{Yuan2015,Yuan2022,Yuan2025} satellite, launched on January 9, 2024, is enabling a transformation of the field, discovering $\approx$120 new extragalactic FXTs in the first $\sim$1.5~years of operations.
EP contains the Wide-field X-ray Telescope (EP-WXT) with a field of view (FoV) of 3600 square degrees, the first wide-field imaging survey instrument operating in the soft X-ray band (0.5--4 keV). It is specifically designed for precise localization and rapid reporting of FXTs.
The delay between discovery and multi-wavelength follow-up observations is reduced even by up to a factor of $\sim10^4$ in time with respect to the FXTs discovered by \emph{Chandra} or \emph{XMM-Newton}.
In addition, in several cases, the Follow-up X-ray Telescope onboard EP (EP-FXT) subsequently provides an accurate X-ray position to a few tens of arcseconds on timescales of order an hour or even less when the EP-WXT triggers an onboard automatic slew to the EP-FXT instrument, and for a subsample of these optical/near-infrared (NIR) counterparts have been identified.

Recently, the fast X-ray transient EP~240414a was discovered by the EP satellite \citep{Srivastav2025,Bright2025,van_Dalen2024,Sun2024}. It was found at a redshift of $ z\approx0.4$ and associated with a massive host galaxy, located at a large projected physical offset of $\approx$26~kpc \citep{van_Dalen2024,Srivastav2025}. According to the Amati relation, its energetics are consistent with those of low-luminosity GRBs \citep{Sun2024}. The optical light curve showed an initially slow decline \citep[inconsistent with afterglow emission;][]{van_Dalen2024}, followed by a re-brightening to $M_r \approx -21$~mag at day $\sim$2 in rest-frame (rising from the interaction with the stellar envelope and a dense circumstellar medium), consistent with the behavior of LFBOTs. Furthermore, at later epochs (after $\sim$15 days, rest-frame), a third component emerged, peaking at $M_r \approx -19.5$~mag, coinciding with the spectroscopic confirmation of a Type Ic-BL supernova. This spectroscopic identification confirms a massive star as the progenitor of the transient, making EP~240414a the first object to establish a direct link between the progenitors of long-duration GRBs, FXTs, and (potentially) LFBOTs \citep{van_Dalen2024}.

In this paper, we study in detail the EP transient EP~241021a. EP~241021a is a newly discovered FXT detected on 2024 October 21 by the EP satellite through its EP-WXT instrument \citep{Hu2024GCN}. The transient exhibited a luminous soft X-ray flash with a duration of approximately $\sim$100 seconds and a peak 0.5–4 keV luminosity of $\sim10^{48}$~erg~s$^{-1}$ \citep{Shu2025} at a redshift of $z\approx0.75$ \citep{Pugliese2024GCN}. No gamma-ray counterpart was detected for this event \citep{Burns2024GCN}. Recently, this transient has been studied in detail by different groups using radio, optical, NIR, and X-ray data \citep[e.g.,][]{Busmann2025,Yadav2025,Gianfagna2025,Shu2025,Wu2025a},which suggested different interpretations.
The optical and NIR light curves show an initially slow decay followed by a steeper rise after a few days, peaking at an absolute magnitude of $M_r\approx-22$~mag \citep{Busmann2025}. The multi-wavelength spectral energy distribution (SED, until day $\sim18$ after the detection) is consistent with a non-thermal origin \citep[$r-J\approx1$ and $r-z\approx0.4$;][]{Busmann2025,Shu2025}. Some authors suggest an association with a mildly relativistic outflow from a low luminosity GRB \citep{Busmann2025}, or the interaction of a complex/multi-component jet (from a highly relativistic narrow cone to much a broader and mildly relativistic cocoon) with the environment of the pre-existing progenitor \citep{Gianfagna2025,Yadav2025}. Alternative scenarios considered include a catastrophic collapse/merger of a compact star system leading to the formation of a millisecond magnetar \citep{Wu2025a} or a jetted tidal disruption event (TDE) involving three jet ejections \citep{Shu2025}. 

Here, we report on the multi-wavelength dataset of the FXT EP~241021a \citep{Hu2024GCN} that our team gathered. We present extensive imaging and spectroscopic observations spanning from $\sim$1.5 to 300 days after the EP-WXT trigger. We analyze our dataset in combination with the data reported in the literature \citep{Busmann2025,Yadav2025,Gianfagna2025,Shu2025,Wu2025a}. 
In Section \S\ref{sec:observations}, we introduce the multi-wavelength observations used in this work. In Section \S\ref{sec:results}, we describe the most significant results obtained by analyzing the multi-wavelength data. 
Section \S\ref{sec:pheno} shows a phenomenological discussion on the nature of EP~241021a, event rate density of similar transients (\S\ref{sec:rates}), and host galaxy properties (\S\ref{sec:host}). Meanwhile, our multi-wavelength light curve modeling is presented in \S\ref{sec:modeling}.
Finally, in Section \S\ref{sec:conclusion} we summarize our main results and interpretation. Throughout we provide magnitudes in the AB system, and for near-infrared (NIR) magnitudes calibrated to 2MASS, which is in the Vega system, we use the Vega to AB conversions  $J_{\mathrm{AB}} = J_{\mathrm{VEGA}} + 0.91$, $H_{\mathrm{}{AB}} = H_{\mathrm{VEGA}} + 1.39$, $K_{s,\mathrm{AB}} = K_{s,\mathrm{VEGA}} + 1.85$ \citep{Blanton2007}. We assume a $\Lambda$CDM cosmology with $H_0=67.66$ km s$^{-1}$ Mpc$^{-1}$ and $\Omega_{\Lambda}=0.69$.


\begin{figure*}
    \centering
\includegraphics[scale=0.81]{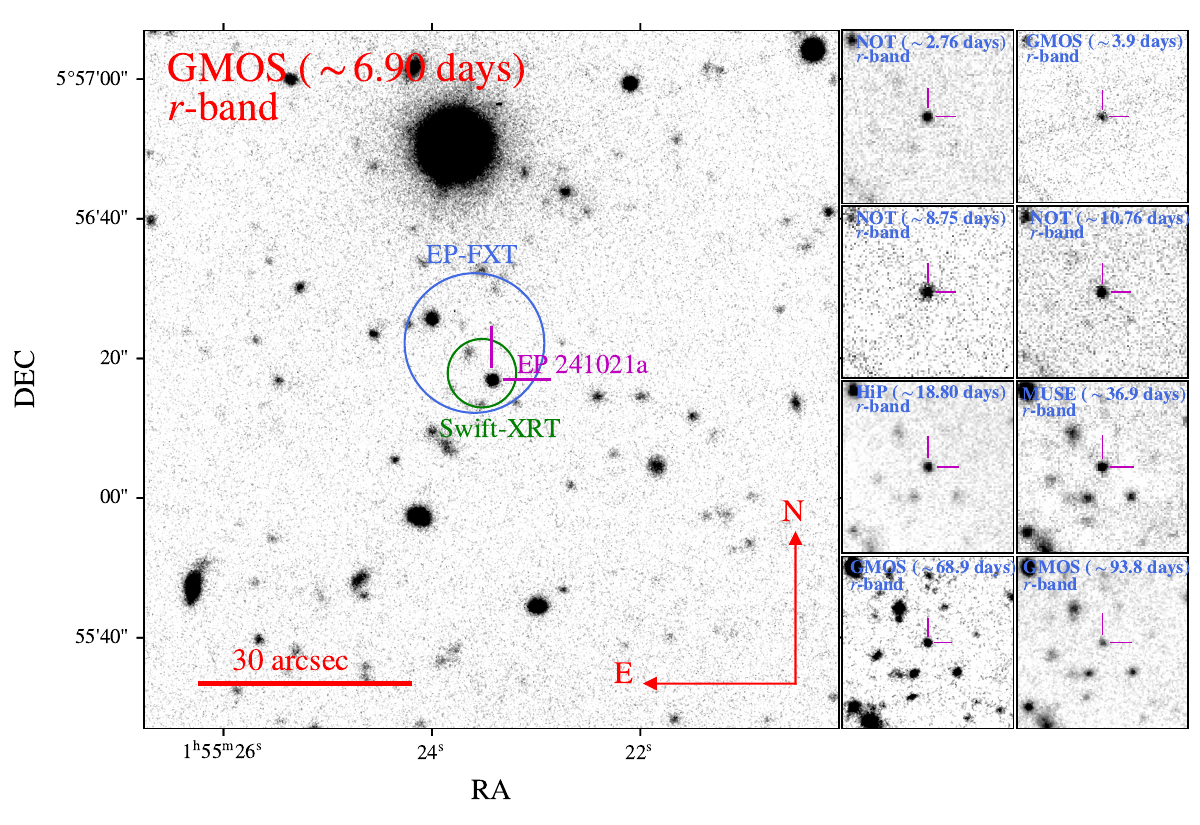}
    \vspace{-0.4cm}
    \caption{
    $r$-band image of the X-ray transient EP~241021a; the position of the event is marked by magenta lines. The large panel is made from the GMOS observation of EP~241021a at day $\sim6.9$. The blue (10\arcsec radius) and green (4.9\arcsec radius) circles show the location of the transient reported by the EP Follow-up X-ray Telescope and \emph{Swift}-XRT, respectively. The lateral panels display eight snapshots of the photometric evolution in $r$-band of the transient, from day $\sim2.76$ to $93.8$, taken by different telescopes.
    }
    \label{fig:time_line}
\end{figure*}

\section{Observations}\label{sec:observations}

The EP-WXT detected the X-ray transient EP~241021a on 2024-10-21 05:07:56 UTC \citep{Hu2024GCN}. The source had a peak flux of $\sim1\times10^{-9}$~erg~s$^{-1}$~cm$^{-2}$ in the 0.5-4 keV band, while its average spectrum can be fitted with an absorbed power law with a photon index of $\Gamma=1.5\pm1.2$ and a fixed Galactic column density of $N_{\rm H,Gal}=5\times10^{20}$~cm$^{-2}$, at an average unabsorbed 0.5-4 keV flux of $3.3_{-1.6}^{+4.8}\times10^{-10}$~erg~s$^{-1}$~cm$^{-2}$ at a 90\% confidence level \citep{Hu2024GCN}.

Observations taken by the Nordic Optical Telescope \citep[NOT;][]{Andersen1992}, starting at $\sim$1.77~days after the X-ray trigger, revealed an uncatalogued optical source within the EP-WXT uncertainty position, at coordinates
RA$_{\rm J2000.0}=$01$^{\rm h}$55$^{\rm m}$23$^{\rm s}$.41,  
Dec$_{\rm J2000.0}=$+05$^\circ$56$\arcmin$18$\arcsec$.01 (with uncertainty of 0.5~arcsec; although no confidence interval was given by \citealp{Fu2024GCN}), with a magnitude of $m_z=21.6\pm0.11$ and $m_r=21.95\pm0.06$~mag \citep{Fu2024GCN}. Because of the temporal and spatial coincidence, this source was suggested to be the optical counterpart of EP~241021a. This was confirmed through multiple independent optical observations taken by the Thai Robotic Telescope network located at Fresno, California, U.S.~(TRT-SRO), 
the 0.76-m Katzman Automatic Imaging Telescope (KAIT) located at Lick Observatory, 
and the Liverpool Telescope (LT) located at the Observatorio del Roque de los Muchachos, La Palma, Spain \citep{Fu2024GCNb,Li2024GCNa,Ror2024GCN,Li2024GCNa} who reported variability in the optical brightness. 
The counterpart was further confirmed by an X-ray observation taken by the EP-FXT ($\sim$36.5 hours after the EP-WXT detection), which refined the X-ray position to be consistent with the proposed optical counterpart \citep{Wang2024GCN}.

Our optical spectroscopic observations began using the FOcal Reducer and low dispersion Spectrograph \citep[FORS;][]{Appenzeller1998} mounted on the Very Large Telescope (VLT) $\sim$3~days after the EP-WXT trigger. The identification of a strong emission line [O II] at 6519~\AA~and the doublet absorption Mg II at 4897~\AA, suggests a counterpart redshift of $z\approx0.75$ \citep{Pugliese2024GCN}. Subsequent independent spectroscopic observations using the Gran Telescopio Canarias \citep{Perez2024GCN}, Keck \citep{Zheng2024GCN}, and Gemini-South (this work) telescopes confirm the redshift of the transient.

Several ground- and space-based telescopes observed the transient EP~241021a over the period ranging from $\sim$1.7 to 100~days after the X-ray trigger. Figure~\ref{fig:time_line} shows the environment as well as eight imaging snapshots highlighting the photometric evolution of the transient EP~241021a in the $r$-band from day $\sim3.76$ to $93.8$. From \S\ref{sec:NOT} to \S\ref{sec:CXO}, we introduce the multi-wavelength observations used in this work (chronologically sorted), as well as a description of the data reduction methods. The optical and NIR photometric light curves of EP~241021a are shown in Fig.~\ref{fig:lc}, while the optical spectra of the transient are in Fig.~\ref{fig:spectra}. All the photometry is aperture photometry and calibrated to Pan-STARRS, Legacy, or 2MASS stars in the field.

The photometric results, derived in this work, are provided in Table~\ref{tab:photometry}. Moreover, throughout this manuscript, we used multi-wavelength data taken from GCNs and the literature (see Tables~\ref{tab:photometry_gcn}, \ref{tab:x-rays}, \ref{tab:radio}, including references to the original works).

\subsection{Nordic Optical Telescope: ALFOSC camera}\label{sec:NOT}


We also observed the target using the Andalucia Faint Object Spectrograph and Camera (ALFOSC) mounted on the Nordic Optical Telescope (NOT) at the Roque de los Muchachos observatory (Canary Islands, Spain) on multiple epochs (from $\sim$1.7 to 15.8~days after the trigger) in the $g$, $r$, and $z$-bands  (program NOT 70-301, PI Jonker). We use a standard dithering pattern for all observations, while the images were reduced with standard reduction steps employing IRAF \citep{Tody1986}. 

\subsection{Gran Telescopio Canarias (GTC)}

The GTC at the Roque de los Muchachos observatory (Canary Islands, Spain) observed the source using the OSIRIS+, HiPERCAM, and EMIR instruments on several nights (program GTC1-24ITP, PI Jonker).

\subsubsection{OSIRIS+ spectroscopy}

We obtained two spectroscopic epochs of the optical counterpart of EP~241021a with the upgraded Optical System for Imaging and low-intermediate-Resolution Integrated Spectroscopy (OSIRIS+) instrument \citep{Cepa2000}. 
The first spectrum of the target 
was obtained on 31~October (i.e., $\sim$9.9~days, after the X-ray trigger) using an exposure time of 3$\times$1800~sec, the grism R500R (covering $\approx$5000-9000~\AA), and a 1\arcsec slit width placed at a position angle North-South. 
The second epoch 
was taken on 9~November (i.e., 19.8~days after the X-ray trigger) using an exposure time of 3$\times$1800~sec, the grism R300B (covering $\approx$4000-8500~\AA), and a 1\arcsec slit width placed aligned East--West. 
The data of both spectroscopic epochs were reduced in the same way: the data were corrected for bias and flatfield, and the source spectrum was extracted using standard \texttt{PyRAF} tasks. Cosmic rays were removed using \texttt{lacosmic} \citep{van_Dokkum2001}. The wavelength calibration was done using daytime arc-lamp observations using \texttt{MOLLY}. The flux calibration was done through observations of a standard star taken on the same night.

\subsubsection{HiPERCAM photometry}

We obtained photometric data of our target using the High PERformance CAMera \citep[HiPERCAM;][]{Dhillon2021} in the $ugriz$-bands, simultaneously, during five different epochs of 40$\times$30~sec exposure time: 24~October (i.e., $\sim$~2.8 days after the trigger), 29~October (i.e., $\sim$~7.9 days after the trigger), 9~November (i.e., $\sim$~18.8 days after the trigger), 22~November (i.e., $\sim$~32.7 days after the trigger), and 1~February 2025 (i.e., $\sim$~103.7 days after the trigger). The obtained data were reduced with the HiPERCAM pipeline \citep{Dhillon2021}. The source is detected in all bands. 
Moreover, to detect the host galaxy of the transient, we observed the transient's sky location using HiPERCAM on 2025-08-21 03:38:10.9 UTC (i.e., $\sim$304 days after the X-ray trigger), when the light of the transient had disappeared, obtaining photometry of the host in the $ugriz$-bands simultaneously, using $120\times30$~sec exposures. The results of the host galaxy detection and photometry are available in \S\ref{sec:host}.

\subsubsection{EMIR photometry}

The target was also observed by the Espectr\'ografo Multiobjeto Infra-Rojo \citep[EMIR;][]{Garzon2022} in the NIR range on 10~November (i.e., $\sim$20.8~days after the X-ray trigger). The target was observed in the $H$ band for 1280~s, where the target was detected. The EMIR data were reduced using \texttt{PyEMIR} \citep{Pascual2010,Cardiel2019} and standard IRAF tasks \citep{Tody1986} to perform flat-field correction and stack individual frames. 

\subsection{Very Large Telescope (VLT)}

\subsubsection{FORS2 spectroscopy}

The FORS2 instrument \citep{Appenzeller1998} was used to obtain spectroscopy (program ESO 114.27PZ, PI Tanvir) using the 300V grism (covering $\approx$3300--6600~\AA), an exposure time of $4\times1200$~sec, and a $1\arcsec$ slit width on 24 October ($\sim$3 days after the trigger).
We reduced the spectroscopic observations using the FORS2 \texttt{esoreflex} pipeline \citep{ESO2015,Freudling2013}. We flux-calibrated the spectrum employing observations of the spectrophotometric standard obtained with the same setup.

\subsubsection{Multi Unit Spectroscopic Explorer (MUSE)}

We obtained an observation of EP~241021a with the MUSE instrument \citep[covering $\approx$4800--9300~\AA;][]{Bacon2010} on 27 November, i.e., $\sim$37~days since the trigger (program ESO 111.259Q, PI Jonker), and an exposure time of $4\times700$~sec. The data was reduced using the MUSE \texttt{esoreflex} pipeline \citep{ESO2015,Weilbacher2020}, and additional sky subtraction was performed with the Zurich Atmosphere Purge (ZAP) package \citep{Soto2016}.
The reduced MUSE cube was visually inspected using the viewer package \texttt{QFitsView} \citep{Ott2012}.
We created broad-band images in the  $r$-, $i$-, and $z$-band using the Python package \texttt{PyMUSE} \citep{Pessa2018}. For the spectral extraction, we also use \texttt{PyMUSE}, and a circular aperture of 8 pixels (i.e., 1.6$^{\prime\prime}$ aperture) because of the point-like nature of the target. For combining pixels within the spectral extraction aperture, we use the mode ``White Weighted Mean''. In this mode, the spectrum from an aperture is a weighted sum of the spaxels by a brightness profile obtained from the white-light image.



\subsection{Gemini South: The Gemini Multi-Object Spectrographs (GMOS)}

The target was observed using the GMOS instrument \citep{Hook2004,Gimeno2016}, mounted on the Gemini-South (GS) telescope at Cerro Pach\'on, Chile. GMOS was operated in imaging mode during several epochs between $\sim$7 to 95 days after the X-ray trigger (using programs GS-2024B-Q-131 and GS-2024B-FT-112, Bauer PI). 
We reduced GMOS imaging data using the {\sc DRAGONS} pipeline \citep{Labrie2023a,Labrie2023b} and followed standard recipes for the imaging and spectroscopic observations. 
In addition to the photometric data, we obtained a spectrum of the optical counterpart of EP~241021a at three different epochs, namely, on 25~October at $\sim$3.9~days, on 28~October at $\sim$7~days, and on 31~October at $\sim$10.1~days, with a total on-target exposure time of $4\times900$, $4\times1500$, and $4\times1200$~sec, respectively. Our spectroscopic observations were all carried out using the GMOS-S instrument (program GS-2024B-Q-131, Bauer PI), the R400 grating, a 1\arcsec slit width, and with the slit oriented at the parallactic angle. The spectra cover the wavelength range $\approx$4200–9100 \AA. 
Moreover, the reduced spectra were flux calibrated using a standard star taken on previous nights of the spectroscopic observations. During the three spectroscopic epochs, there are a number of emission lines evident in the spectrum that can be used to estimate the redshift of the system.

\subsection{Keck Telescope: MOSFIRE}


NIR imaging observations, using the Multi-Object Spectrograph for Infrared Exploration \citep[MOSFIRE;][]{McLean2012} at the Keck I telescope (Mauna Kea, Hawaii) were taken using the broadband filters $J$, $H$, and $K_s$ on 25 October (i.e., $\sim$4.2 days after the X-ray trigger), with exposures of 5$\times$60, 6$\times$30, and 9$\times$60~sec, respectively. The data were reduced with the software package \texttt{THELI} version 3.1.3 \citep{Erben2005,Schirmer2013}.
Clear detections of the counterpart were obtained in the three NIR bands.



\subsection{Southern Astrophysical Research (SOAR) Telescope: the Goodman Red and Blue Camera}


The Goodman Blue and Red Camera instruments mounted on the 4.1-meter SOAR Telescope located on Cerro Pach\'on, Chile \citep{Clemens2004} were used to observe the optical counterpart of EP~241021a, during four different epochs (program SOAR2024B-016, PI Bauer). Data was obtained on 27~October at $\sim$5.9~days after the EP-WXT trigger (using $i$ band), on 31~October at $\sim$9.9~days since the trigger (using $r$- and $i$-bands), on 3~November at $\sim$12.8~days since the trigger (using $g$- and $r$-bands), and on 18~November at $\sim$27.9~days after the trigger (using $r$- and $z$-bands). The data were bias subtracted and flat-field corrected adopting standard \texttt{PyRAF} tasks \citep{Pyraf2012}, and cosmic rays were removed using the \texttt{LACosmic} task \citep{van_Dokkum2001}. The world coordinate system information is implemented using \texttt{astrometry.net} codes \citep{Lang2010}. The counterpart of EP~241021a was clearly detected during the first three epochs and marginally detected during the last one. 

\begin{figure}
    \centering
    \includegraphics[scale=0.5]{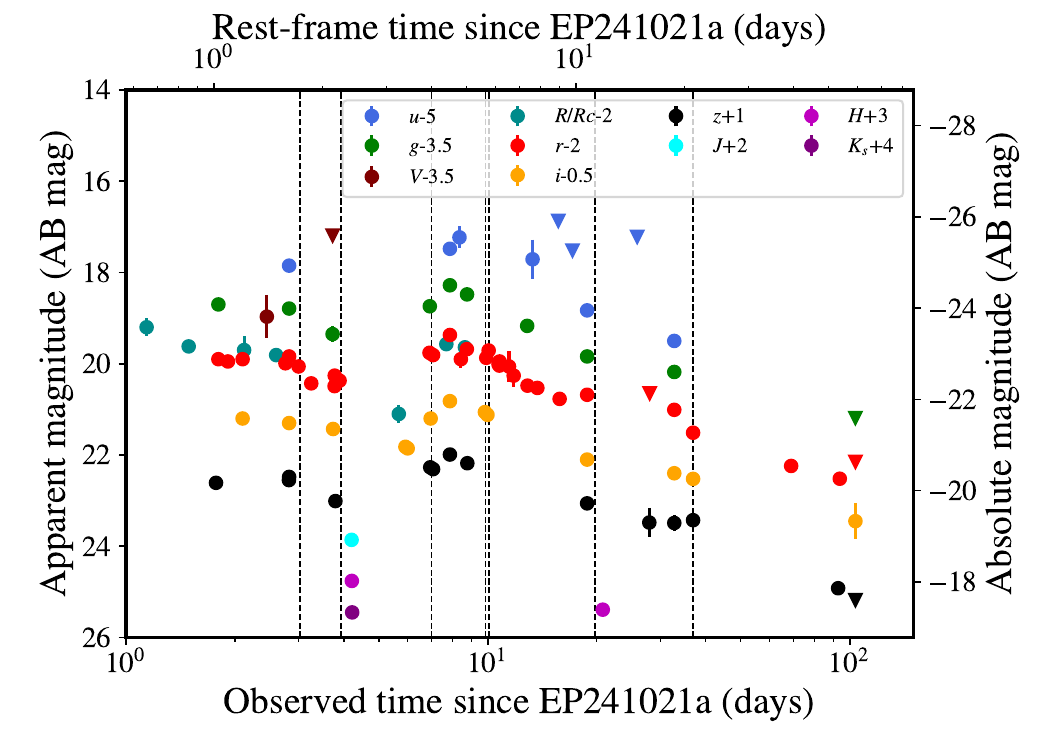}
    \vspace{-0.5cm}
    \caption{Apparent (left Y-axis) and absolute (right Y-axis) magnitude light curves of FXT EP~241021a in the observer- (bottom X-axis), and rest frame (top X-axis) in several optical and NIR bands. The times at which our spectroscopic observations were taken are marked by the vertical black dashed lines.
    }
    \label{fig:lc}
\end{figure}

\begin{figure*}
    \centering
    \includegraphics[scale=0.65]{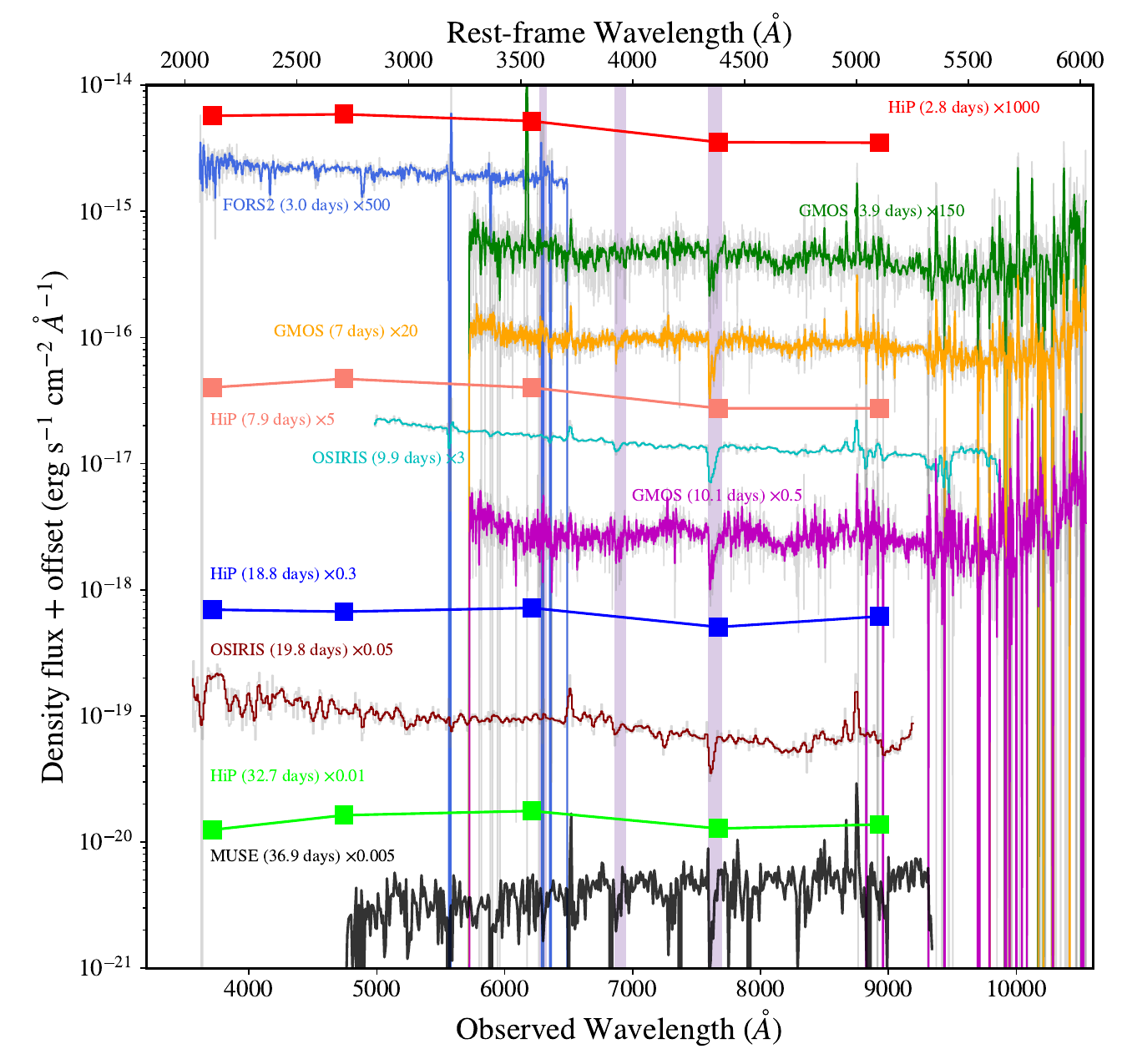}
    \vspace{-0.4cm}
    \caption{Evolution of the optical spectra of the counterpart of FXT EP~241021a in the observer (bottom X-axis) and rest (top X-axis) frame. The spectrum obtained first is shown at the top, and the last one is at the bottom (the times in brackets next to the spectra indicate the days after the WXT trigger, in the observer frame). In addition, we also show the flux density measured using our four GTC/HiPERCAM observations ($ugriz$ bands). Vertical regions corresponding to telluric absorption are shaded.}
    \label{fig:spectra}
\end{figure*}

\subsection{Neil Gehrels Swift Observatory}

The \textit{Swift} satellite \citep{Gehrels2004} observed the location of EP~241021a on seven different epochs, with the X-ray telescope (XRT) and the UltraViolet/Optical Telescope (UVOT). The data span the time range between $\sim$4.5 and 25.9 days after the trigger, for a total exposure time of about $\sim$17.3~ks. 

\subsubsection{X-ray telescope (XRT)}

From the data products generator online tool provided by the University of Leicester \citep{Evans2007,Evans2009},\footnote{\hyperlink{https://www.swift.ac.uk/user_objects/}{https://www.swift.ac.uk/user\_objects/}} the X-ray afterglow of EP~241021a was detected, localized at 
RA$_{\rm J2000}=$01$^{\rm h}$55$^{\rm m}$23$^{\rm s}$.52,  
Dec$_{\rm J2000}=$+05$^\circ$56$\arcmin$17$\arcsec$.8
with uncertainty of 4.9\arcsec at 90\% confidence level (using HEASOFT v6.35.2).
We extracted the source count rate using the command line interface \texttt{Xselect} \citep{ingham2006} and HEASOFT v6.32, using a region of 40$^{\prime\prime}$ radius centered at the position of the optical counterpart \citep{Fu2024GCN}, while the background counts were extracted from a larger region (60$^{\prime\prime}$) near the source.
The corresponding response files for spectral fits were created with the latest calibration available for \emph{Swift} data (updated version 2024 May). Unfortunately, a detailed spectral analysis per individual observation is not possible due to the low number of X-ray counts.
Moderate detections (i.e., S/N$\gtrsim$3) were achieved only at two epochs ($\sim$8.3 and 25.8~days after the trigger). Our results are consistent with the count-rate light curve retrieved by the data products generator tool. Combining and fitted all epochs using a power-law model \citep[and a fixed Galactic hydrogen column density of $5\times10^{20}$~cm$^{-2}$;][]{Willingale2013,Kalberla2005,Kalberla2015} yields a photon index of $\Gamma=2.3_{-0.7}^{+1.3}$ and an unabsorbed energy conversion factor of $3.1\times10^{-11}$~erg~cm$^{-2}$~cnt$^{-1}$ (used to transform the count rate to flux units). 
Table~\ref{tab:x-rays} shows the derived unabsorbed fluxes of the target.

\subsubsection{UltraViolet/Optical Telescope (UVOT)}

Furthermore, we analyzed the images obtained by the \emph{Swift}-UVOT in the optical and UV filters. \emph{Swift}-UVOT observed the target with filters $V$, $U$, and $UVM2$ during seven different epochs. 
The filters $V$ and $UVM2$ were used only during two epochs ($\sim$3.7 and 4.5 days) and $U$-band during the other observations. 
To measure photometry, we used the script \texttt{uvotsource}, which performs aperture photometry on a single source in a UVOT sky image. We use a circular aperture of $4^{\prime\prime}$ (because of the faintness of the target). The package is part of the \texttt{HEASOFT} software package \citep{Blackburn1995}. 
During days $\sim$8.3 and 13.3, the target was significantly (S/N$\approx$5) and marginally (S/N$\approx$3) detected, respectively, using filter $u$.

\subsection{\emph{Chandra} X-ray Observatory} \label{sec:CXO}

The {\emph Chandra} satellite observed the location of EP~241021a on 4 November ($\sim$14~days after the trigger) for a total exposure time of 10~ks (Director's Discretionary Time; PI Jonker). The source position was placed at the S3 CCD of the ACIS-S detector array using the very faint mode for read-out \citep{Garmire1997}. We analyze the data using the package {\sc ciao}~version 4.16 \citep[Chandra Interactive Analysis of Observations;][]{Fruscione2006}, reprocessing the events using the task {\sc acis\_reprocess\_events}, taking the {\sc very faint} data telemetry mode into account.
Using {\sc wavdetect} \citep{Freeman2002} to detect sources, we detect X-ray emission spatially coincident with the location of the optical counterpart of EP~241021a. The spectrum was extracted via the {\sc ciao} package \texttt{specextract}. To obtain the X-ray spectral parameters, we used the spectral fitting program \texttt{XSPEC} v12.14 \citep{Arnaud1996} and considered the Cash statistic \citep{Cash1979} to account for the low number of counts of the target (26 counts inside an circular aperture region with a radius of $1.5^{\prime\prime}$).
We fit an absorbed power-law model (\texttt{tbabs*pow} model in \texttt{XSPEC}) to the spectrum binned to have at least one photon per bin, where the \texttt{tbabs} model describes the Galactic absorption. For the extraction of the X-ray spectral parameters, we fixed the Galactic hydrogen column density to $5\times$10$^{20}$~cm$^{-2}$ \citep{Kalberla2005, Kalberla2015}. The absorbed fluxes are computed using the standard \texttt{XSPEC} tasks, while the unabsorbed fluxes are computed by the \texttt{XSPEC} convolution model \texttt{cflux} (see Table~\ref{tab:x-rays}).
The best-fit power law index is $\Gamma=1.3\pm0.7$ (at 90\% confidence level) and the best-fit source absorbed flux $F_X$=$\left(6.5^{+3.1}_{-2.0}\right)\times10^{-14}$~erg~cm$^{-2}$~s$^{-1}$ at 0.3-10~keV for a C-stat of 21.76 for 21 degrees of freedom (dof).

\begin{figure}
    \centering
    \includegraphics[scale=0.65]{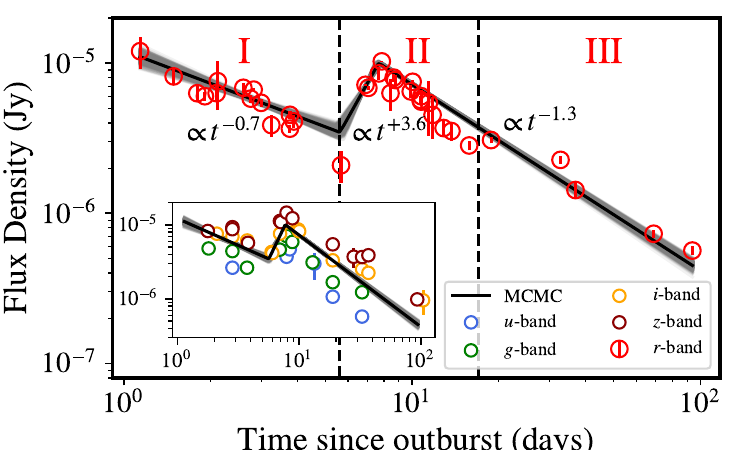}
    \vspace{-0.1cm}
    \caption{$r$-band light curve fit of the FXT EP~241021a using a smooth triple-power-law model (black line). The source flux decays initially following approximately a power-law,
    then it re-brightens around $t_1=5.6\pm0.2$~days
    reaching a peak flux at $t_2=7.7\pm0.1$~days, and it subsequently decays, which can be described approximately by another power-law.
    The Roman numbers at the top depict the three epochs of the light curve: 
    (I) initial decay at $t\lesssim6$~days; 
    (II) re-brightening and a peak emission followed by a decay at $6\lesssim t\lesssim18$~days; and 
    (III) a re-brightening followed by power-law decay starting at $t\approx18$~days.
    We can expect a similar light curve behavior in other bands; however, the sampling cadence is not identical for all of them.
    Moreover, the inside figure depicts a comparison between the smooth triple-power-law model and $ugiz$ data. The units in the inside plot are the same as the large plot.
}
    \label{fig:lc_fit}
\end{figure}


\section{Results} \label{sec:results}

\subsection{Redshift and Energetics}

The redshift of $z=0.75$ was determined from emission lines in the 2d spectra of the target by multiple groups with different instruments \citep{Pugliese2024GCN,Perez2024GCN,Zheng2024GCN}. To refine the reported redshift, we fitted the emission lines with the highest S/N (H$\beta$, [O II] $\lambda$3727, [O III]$\lambda\lambda$4959,5007\AA) using multiple Gaussian functions, to the GMOS and OSIRIS+ spectroscopic epochs, and obtained the best-fitting central wavelengths and their associated errors. Table~\ref{tab:redshifts} shows the derived redshift per spectroscopic epoch, with a mean value of $\overline{z}_{\rm spec}=0.7485$. Moreover, we detect absorption lines of the Mg II doublet $\lambda\lambda$2796, 2803 \AA~in the FORS2 spectrum at centroid wavelengths consistent with the redshift derived from the emission line fits. We interpret the emission lines as due to ionized gas in the host galaxy.
At this redshift, the transient has a peak luminosity of $L_{\rm X,peak}\approx2\times10^{48}$~erg~s$^{-1}$ in the 0.5--4~keV energy band.

\subsection{Optical Light curve} \label{sec:LC}

The optical and NIR light curves are shown in Fig.~\ref{fig:lc}.
The earliest observation of the target (using the 0.76-m Katzman Automatic Imaging Telescope; KAIT) yields an apparent magnitude of $m_R=21.2\pm0.2$~mag \citep{Zheng2024GCNa}, i.e., an absolute magnitude of $M_R\sim-21.6$~mag at $\sim$1.14~d after the EP-WXT detection. We decided to add this photometric point (and others from literature, see Table~\ref{tab:photometry_gcn}) to the $r$-band light curve\footnote{To check the consistency of the photometric data at $r$ and $R$ bands, we compute the ratio of the integrated spectra convolved by the transmission functions of both filters. We obtain that the ratio between the $r$ and $R$-band fluxes is $\sim$1.04, i.e., the difference between the flux in both bands is of the order of 4\%.}. 
Looking at the optical light curves, we can identify different trends during its evolution.\footnote{To describe the light curve quantitatively, we fit a three-power-law model to the $r$-band light curve (the best sampled) in different phases, as $F\propto t^\alpha$.}
Figure~\ref{fig:lc_fit} depicts the fitting of the $r$-band light curve, which is the best sampled out of the light curves, considering a smooth triple-power-law model. In addition, in the inset in Fig.~\ref{fig:lc_fit}, we show the comparison between the best-fit model to the $r$-band light curve and the $ugiz$ light curves to inspect how the multi-wavelength light curve behaves with respect to that of the $r$ band; it is clear that the transient becomes redder with time. Overall, the optical light curve shows two peaks: the first around $\sim$1~d (no observations prior to this epoch have been reported), and a second one at $7.7\pm0.1$~d (see Fig.~\ref{fig:lc_fit}). During the first $\sim$6~days, the light curves in all the optical bands decrease as a power law function, with a power-law index of $\alpha_1=-0.7\pm0.1$, i.e, $\Delta m\approx$1.9~mag over this period. 

After $t\approx$5.6~d, the source re-brightened in all the bands \citep[reported by][]{Quirola2024GCN}, with a steep slope of $\alpha_2=+3.6\pm0.6$. The absolute magnitude reached a peak at $7.7\pm0.1$~d of $M_g\approx-21.0$, $M_r\approx-21.4$, $M_i\approx-21.5$, and $M_z\approx-21.8$~mag ($\Delta m_{griz}\approx1.0-1.7$~mag).

Subsequently, the light curves decrease rapidly. Fitting all data beyond $\approx7.7$~d, we obtain a power-law index of $\alpha_3=-1.3\pm0.1$ (see Fig.~\ref{fig:lc_fit}). Considering only data from $\approx7.7$~d to 16~d, a power-law decay in the $griz$ bands ($\alpha_g=-1.7\pm0.1$, $\alpha_r=-1.7\pm0.2$, $\alpha_i=-1.4\pm0.1$ and $\alpha_z=-1.1\pm0.1$) describes the data well, where the decline from the peak is $\Delta m_{griz}\approx$1.0--1.6~mag over about 8.3~d.
 
In the case of the $r$-band light curve, a mild re-brightening occurred at day $\sim$18, reaching an absolute peak of $M_r\approx-20.1$~mag at day $\sim$19. This behavior is not visible in the other bands, probably because of the lower cadence of our observations in those other bands. However, it also appears in observations of other groups \citep[e.g.,][]{Busmann2025}. The $r$-band light curve declines again after day $\sim$19. The behavior in the $z$-band is different; here the flux remains constant around days $\sim27$ and 37, followed by a decay.


The evolution of the light curve led us to categorize it into three epochs depicted in Figure~\ref{fig:lc_fit} by roman numbers: (I) initial decay until $t\lesssim6$~d; (II) re-brightening and second decay between $6\lesssim t\lesssim18$~d; and (III) re-brightening followed by a final decay beyond $t\gtrsim18$~d. There are no short-time optical flares of the transient in the multiple HiPERCAM epochs\footnote{The dead time between exposures is only 7~msec, which is really what allows a search for "short-time" flares to be done (i.e., there's no chance we missed one due to the CCD readout time being 1-2 mins like a normal camera).} \citep[using the pipeline developed by ][]{Dhillon2021}.

To explore the contribution of the host galaxy to the light of the optical counterpart of EP~241021a, we observed the host galaxy once the light of the transient vanished (see \S\ref{sec:host} for more details). We determined that the host contribution is negligible during the majority of the period covered by our light curve. For instance, during the main peak at $\sim7.7$~days, the transient is $\approx$20, 18, and 13 times brighter than the host in the $grz$ bands, respectively. Meanwhile, at day $\sim32$, the target is brighter by a factor of $\approx$3.5, 4, and 3.5 for the $grz$-bands, respectively. Only at times beyond day $\sim70$, the photometry of the transient becomes similar to its host galaxy. Nevertheless, we correct the observed photometric data by subtracting the contribution of the host galaxy obtained by the best-fit galaxy model (see \S\ref{sec:host} for more details) in the $ugrizJHK_s$ bands.

\begin{figure}
    \centering
    \includegraphics[scale=0.7]{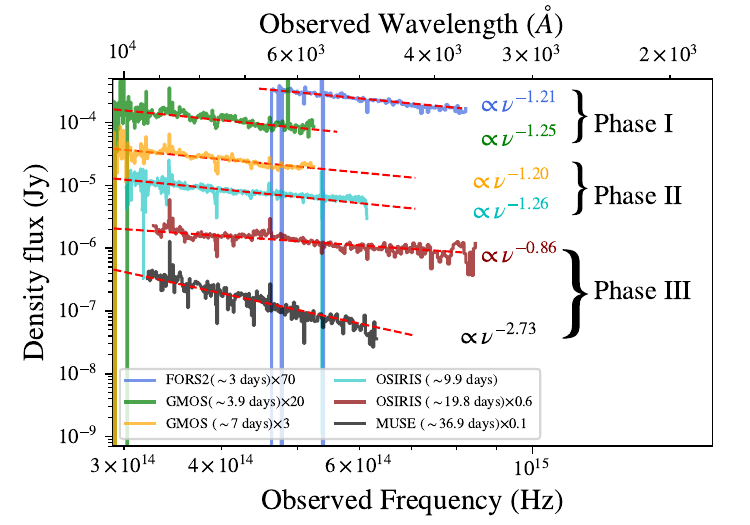}
    \vspace{-0.2cm}
    \caption{
    Spectroscopic evolution of EP~241021a. The red dashed lines depict the power-law model that can describe the continuum of the transient (see \S\ref{sec:spectra} for more details), as well as their best-fit power-law index. For visual purposes, offsets were applied to the spectra (see legend).
    }
    \label{fig:spec_evol}
\end{figure}

\subsection{Spectroscopy}\label{sec:spectra}

We obtained seven optical spectra (see Fig.~\ref{fig:spectra}), covering an (observer) wavelength range from $\sim$3500 to 10,000~\AA. The spectra were obtained over the time interval $\sim3$~d to 37~d after the X-ray trigger. To quantify the spectral evolution, we fit a power-law function (as $F_\nu\propto \nu^{\beta_{\rm spec}}$) to the data. Firstly, the data was binned using \texttt{SpectRes} \citep{Carnall2017} to improve the signal-to-noise ratio (at least to have a S/N$\gtrsim5$ per pixel), and masking sky \citep{Hanuschik2003} and telluric lines, emission lines of the host (i.e., H$\beta$, [O II] $\lambda$3727, [O III]$\lambda\lambda$4959,5007\AA), and correcting for Galactic extinction using the \citet{Calzetti2000} dust model and $E(B-V)=0.045$~mag \citep{Schlafly2011}. 

During Epoch I, we have two spectral epochs taken by FORS2 (day $\sim$3) and GMOS (day $\sim$3.9), covering the range of $\sim$3,500--10,000\AA, with spectral indeces of $\beta_{\rm spec}=-1.21\pm0.04$ ($\chi^2/{\rm dof}=470.58/491$) and $-1.25\pm0.09$ ($\chi^2/{\rm dof}=562.41/334$), respectively (see Fig.~\ref{fig:spec_evol}). Combining both values, the mean value during Epoch I is $\beta^{\rm I}_{\rm spec}=-1.22\pm0.04$.
During Epoch II, we have one spectrum taken during the re-brightening at day $\sim7$ (GMOS), and two more at day $\sim10$ (GMOS \& OSIRIS+) at the decay of the main peak. During days $\sim7$ and $10$, the optical spectral indices are $\beta_{\rm spec}=-1.20\pm0.05$ ($\chi^2/{\rm dof}= 602.56/396$) and $-1.26\pm0.07$, respectively. Combining both values, the mean value during Epoch II is $\beta^{\rm II}_{\rm spec}=-1.23\pm0.05$, which is compatible with that measured during Epoch I.
Overall, until Epoch II, the optical spectra are well-described by a power-law function.
During Epoch~III, we have two spectroscopic observations. On day $\sim19.8$ (OSIRIS+ spectrum), the best power-law fit is $\beta_{\rm spec}=-0.86\pm0.02$ ($\chi^2/{\rm dof}=1179.28/1110$), showing a change in the slope with respect to that measured in Epochs~I and II. 
Furthermore, we also used a broken power law model to explore a potential more complex spectral shape, especially motivated by the flux density peak at $\approx6700$ \AA~(see Fig.~\ref{fig:spec_evol}). We obtained that the broken power-law model ($\chi^2/{\rm dof}=1171.92/1108$) depicts red and blue slopes of $\beta_{\rm spec}^{\rm red}=-0.69\pm0.09$ and $\beta_{\rm spec}^{\rm blue}=-0.92\pm0.04$, respectively, and a break wavelength at $=6750\pm58$~\AA. Although the broken power law model improves the fit, the difference between the models is not statistically significant (the $\chi^2$ decreases by 7.36 for 2 extra degrees of freedom). 
Finally, the last spectrum of EP~241021a during Epochs~III was taken by MUSE at day $\sim36.9$ (see Fig.~\ref{fig:spec_evol}). The spectrum shows an optical spectral index of $\beta_{\rm spec}=-2.73\pm0.05$ ($\chi^2/{\rm dof}=480.71/433$), which is steeper than the slopes measured in the previous phases, demonstrating a spectral evolution of the transient.
Summarizing, the spectral index does not change significantly ($\beta_{\rm spec}\sim-1.2$) until day $\sim20$, and it indicates that non-thermal emission dominates until the decay of the main peak; meanwhile, during Epoch~III, we have a transition around day $\sim$20, interpretating that either the spectral component changed or a new spectral component emerged. 

To investigate if features related to an SN can be detected in any of our spectra, we use the packages \texttt{Gelato} \citep{Harutyunyan2008} and \texttt{SNID} \citep{Blondin2007}. However, we find no robust evidence for the detection of SN absorption lines in our spectra, which might be a consequence of an additional spectral component that dilutes the equivalent width of the expected absorption lines. This, in combination with the redshift of the event, renders the SN features undetectable to our ground-based 8-10~m telescope observations. We remove any contribution from the host galaxy, given that it is faint (see \S\ref{sec:host} for more details).

\begin{figure}
    \centering
    \includegraphics[scale=0.6]{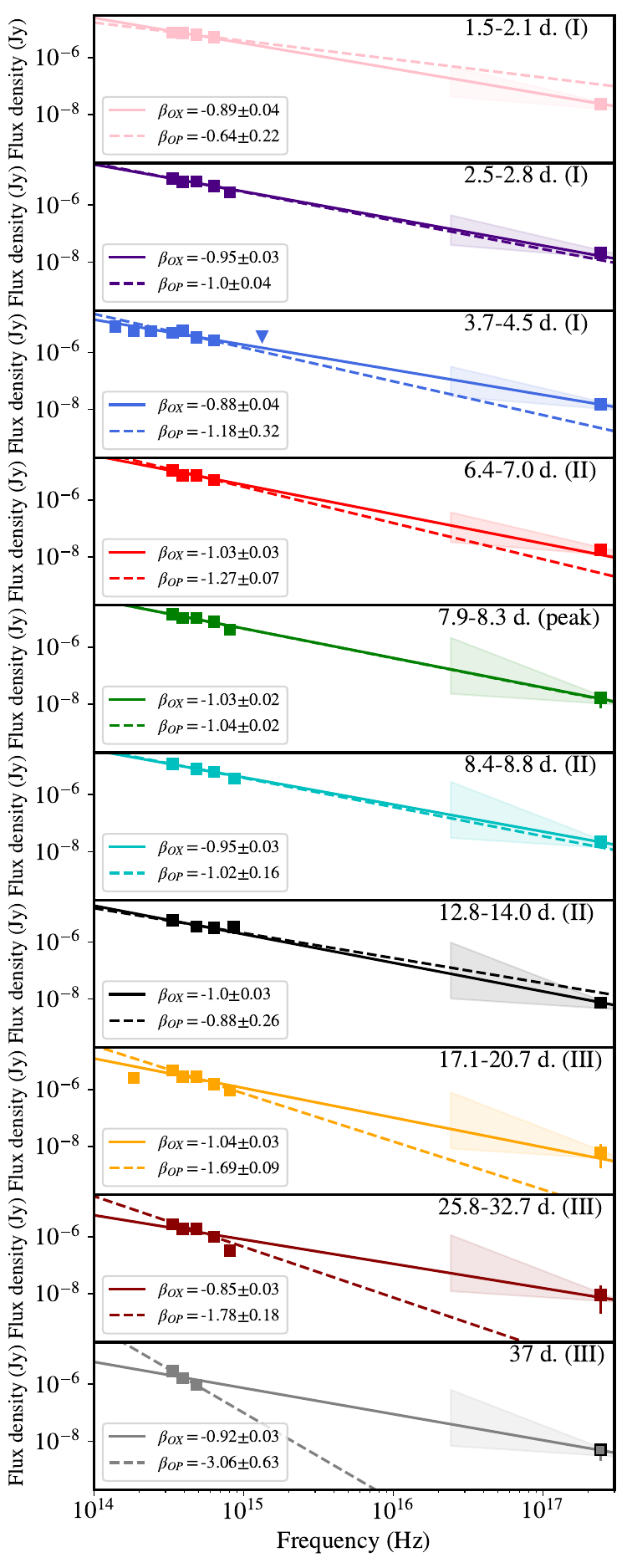}
    \vspace{-0.2cm}
    \caption{Spectral energy distribution (SED) of the transient EP~241021a at different time bins or epochs (Roman numbers in the top-right corner). We fit a power-law function to the optical data ($F_\nu\propto\nu^\beta$), and to the combined optical to X-ray SED. These fits yield the spectral index $\beta_{\rm op}$ and $\beta_{\rm ox}$, respectively. The uncertainty on the spectral index of a power law fit to the X-ray spectrum alone is shown as a shaded region. We use the EP-FXT data from \citet{Shu2025}. The optical and NIR data were corrected for Galactic extinction and for the contribution of the host galaxy.}
    \label{fig:SED}
\end{figure}

\subsection{Optical-to-X-ray Spectral Energy Distribution}

We investigate the evolution of the optical-to-X-ray SED of EP~241021a as a function of time using the available photometric data. First, we correct the photometry for Galactic extinction, i.e., $E(B-V)=0.045$~mag \citep{Schlafly2011} using the laws from \citet{Fitzpatrick1999} and we remove the contribution of the host galaxy (see \S\ref{sec:host}). We consider time bins with at least four photometric points covering NIR, optical, and/or X-ray wavelengths. Figure~\ref{fig:SED} depicts the SED at different time bins. To measure the spectral shape, we fit a power-law function to the optical data (defined by an index $\beta_{\rm op}$) and combine it with the X-ray data (defined by an index $\beta_{\rm ox}$).
Considering only the optical data, we get spectral indices weighted mean at Epoch I, and II of $\beta_{\rm op}^{\rm I}=-1.0\pm0.1$ and $\beta_{\rm op}^{\rm II}=-1.1\pm0.1$, respectively, which both are consistent with our spectroscopic results (see \S\ref{sec:spectra}) at 1-$\sigma$ confidence level. Therefore, until day $\sim$20, the SED at optical wavelength does not change significantly, and indicates non-thermal emission dominates (giving its spectral index). Beyond day $\approx18$, the SED becomes systematically steeper (even reaching a slope of $\approx-3.1$ at day $\sim$37, and confirming the spectroscopic results), which might be interpreted as the emergence of a different spectral component (which is unlikely to be a non-thermal component). The weighted mean of the spectral indices throughout Epoch III is $\beta_{\rm op}^{\rm III}=-1.7\pm0.1$. Moreover, the optical spectral index becomes less consistent with the X-ray photon index at late times (see Fig.~\ref{fig:SED}, last three panels from the top).

The index $\beta_{\rm ox}$ remains practically constant during the whole evolution, with spectral index of $\beta_{\rm ox}=-0.92\pm0.02$, $-1.01\pm0.01$, and $-0.94\pm0.02$ for Epochs I, II, and III, respectively.
The fact that the $\beta_{\rm op}$ and $\beta_{\rm ox}$ are consistent during Epochs I and II suggests that the same spectral component might be responsible for both energy bands until $\sim20$~d. Naturally, the evolution of the optical and X-ray light curves does track each other initially \citep[X-ray data taken by the EP-FXT telescope;][]{Shu2025} until day $\sim20$ \citep{Gianfagna2025,Shu2025}. However, at late times ($t>20$~d), it is difficult to reconcile the steepness of the optical spectral index with the X-ray emission, suggesting different spectral components. Given the faint host galaxy (see \S\ref{sec:host} for more details), we can discard any influence from the host galaxy to explain this change in $\beta_{\rm op}$ during Epoch~III.



\begin{figure}
    \centering
    \includegraphics[scale=0.48]{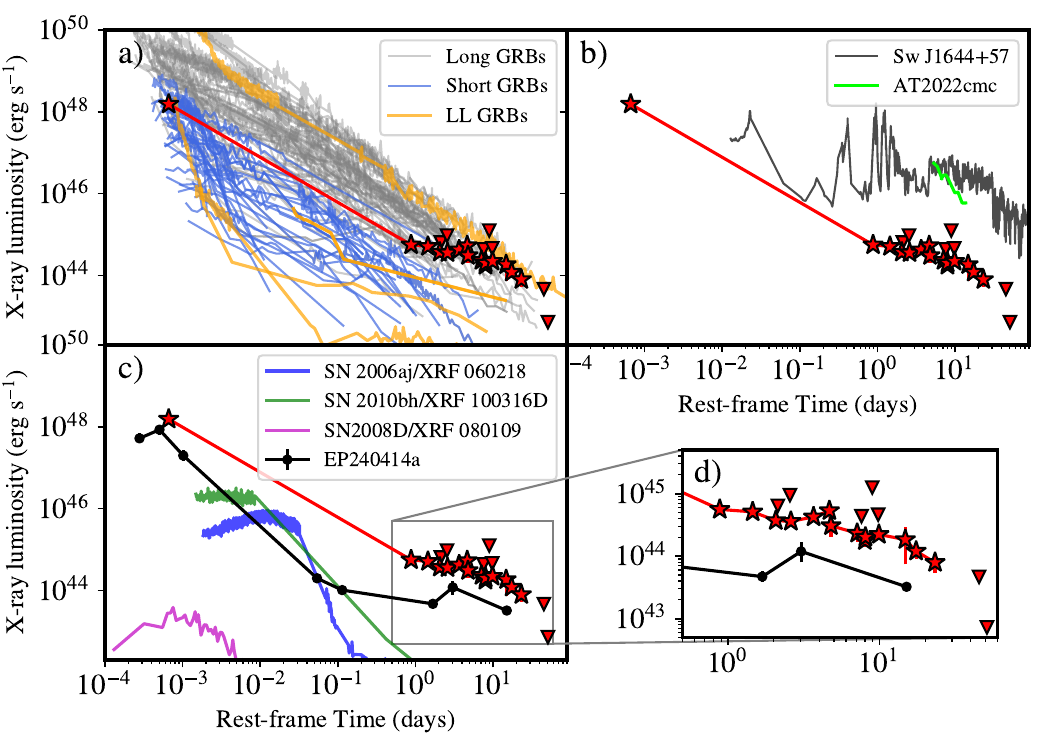}
    \vspace{-0.5cm}
    \caption{X-ray light curve of FXT EP~241021a (red stars) in 0.3--10 keV luminosity units combining EP, \emph{XMM-Newton} \citep{Shu2025}, \emph{Chandra} (this work) and public \emph{Swift}-XRT data. The red triangles depict $3\sigma$ upper limits.
    Each panel depicts a comparison with several individual transients that are overplotted. 
    Panel $a)$: the X-ray afterglow light curves of 64 LGRBs plus 32 SGRBs and low-luminosity GRBs \citep{Evans2007,Evans2009,Bernardini2012,Lu2015};
    Panel $b)$: the relativistically beamed TDEs \emph{Swift}~J1644$+$57 \citep{Bloom2011,Levan2011} and AT2022cmc \citep{Andreoni2022};
    Panel $c)$: the X-ray flashes XRF~060218/SN~2006aj, XRF~100316D/SN~2010bh \citep{Barniol2015,Starling2011,Modjaz2009,Evans2009,Evans2007,Campana2006}, XRF~080109/SN~2008D \citep{Soderberg2008}, and the afterglow of the EP transient EP~240414a \citep{Sun2024,van_Dalen2024}; and
    Panel $d)$: it is a zoom in of the light curve of EP~241021a at late times, with the same units as the other panels.}
    \label{fig:xray_plot}
\end{figure}

\subsection{X-ray light curve}

In Figure~\ref{fig:xray_plot} we show the X-ray light curve of EP~241021a as observed with the EP \citep[EP-WXT and EP-FXT instruments;][]{Shu2025}, \emph{XMM-Newton} \citep{Shu2025}, \emph{Chandra}, and \emph{Swift}-XRT telescopes. As we mentioned before, an approximate peak luminosity of EP~241021a of $L_{\rm X,peak}\approx2\times10^{48}$~erg~s$^{-1}$ is consistent with that of the EP transient EP~240414a (see Fig.~\ref{fig:xray_plot}, panel $c$).
Compared with long GRBs (LGRB), the peak luminosity of EP~241021a falls in the region occupied by low-luminosity (collapsar) long GRBs (see Fig.~\ref{fig:xray_plot}, panel $a$); also, this region is populated by short GRBs (SGRB). The peak luminosity is higher by more than two orders of magnitude than that of the X-ray flashes XRF~060218 and XRF~100316D, and the shock breakout of SN~2008D (see Fig.~\ref{fig:xray_plot}, panel $c$). 
Moreover, the transient at late epochs is brighter than X-ray flashes and even EP~240414a by some orders of magnitude, and remains fainter than the relativistically beamed TDEs Swift J1644+57 \citep{Bloom2002,Levan2011} and AT2022cmc \citep{Andreoni2022} by two orders of magnitude (see Fig.~\ref{fig:xray_plot}, panel $b$).


The X-ray light curve of EP~241021a was fitted using a three-piecewise power-law function by \cite{Shu2025}. The X-ray light curve at $t\gtrsim1$~d shows a plateau phase with a power-law index of $-0.28^{+0.17}_{-0.13}$ during days $\sim1-7$, followed by a decline up to 79 days with an index of $-1.16^{+0.30}_{-1.11}$ \citep[see Fig.~\ref{fig:xray_plot}, panel $d$;][]{Shu2025}; then, the transient becomes undetectable by EP-FXT ($\lesssim2\times10^{-14}$~erg~cm$^{-2}$~$s^{-1}$) and even to \emph{XMM-Newton} ($\lesssim3\times10^{-15}$~erg~cm$^{-2}$~s$^{-1}$) around $\sim89$~days after the trigger, implying a steeper decay index of $-9.64^{+6.92}_{-7.04}$ at day $33.49^{+10.67}_{-9.72}$ \citep{Shu2025}. In addition to these general trends, around day $\sim8$ in the observer's frame, the X-ray transient flux increases ($\approx6\times10^{44}$~erg~s$^{-1}$), which coincides with the main peak in the optical light curve \citep{Gianfagna2025}.
The X-rays at late times fall in a region mostly populated by the afterglow of GRBs, reaching a luminosity of $\approx5\times10^{44}$~erg~s$^{-1}$ (see Fig.~\ref{fig:xray_plot}, panel $a$). 

\begin{figure}
    \centering
    \includegraphics[scale=0.45]{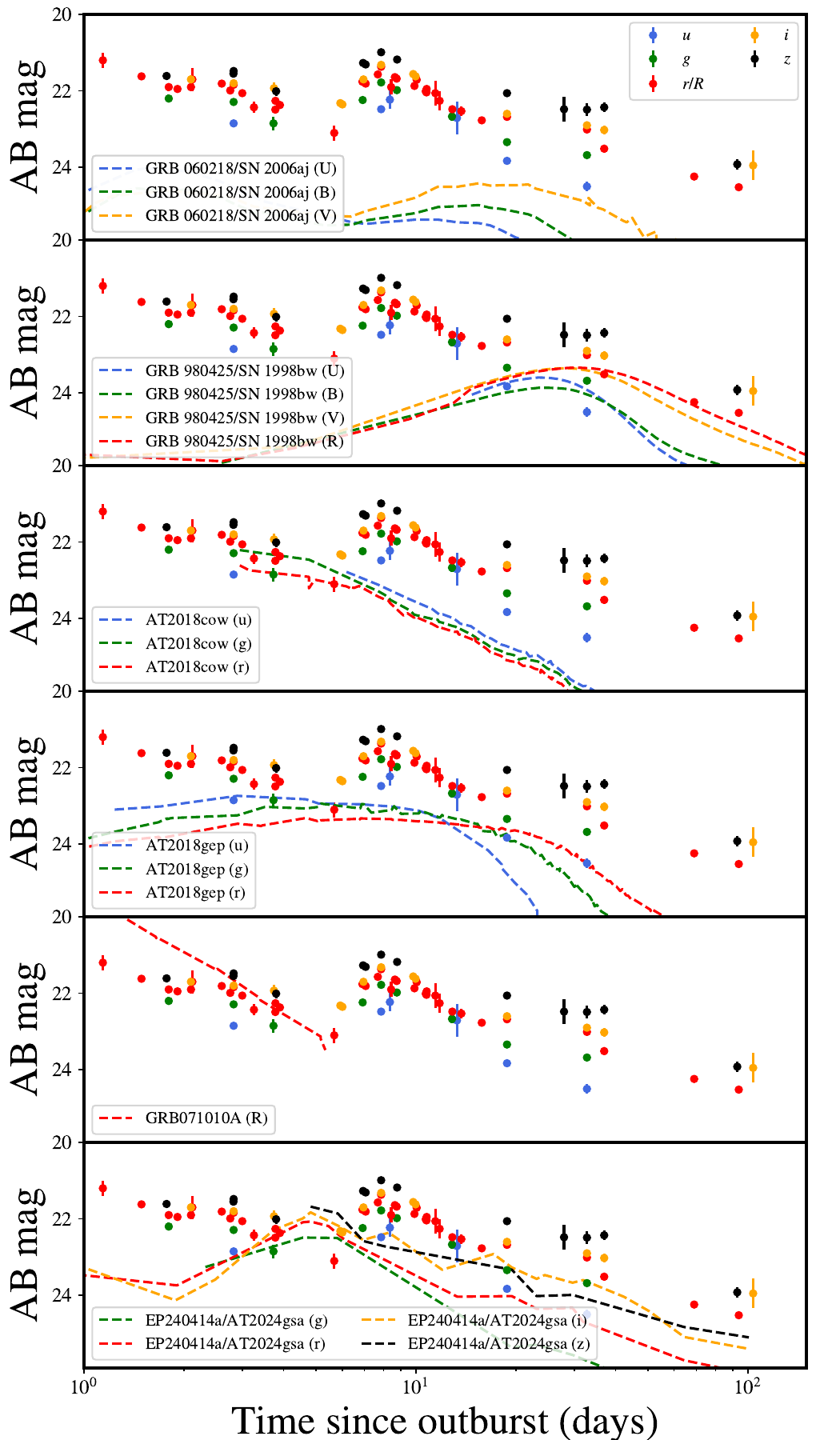}
    \vspace{-0.0cm}
    \caption{Light curve of FXT EP~241021a in the observer frame in different filters ($ugriz$-bands) compared to the light curves of the X-ray flash GRB/XRF 060218 \citep[first panel;][]{Soderberg2006}, the prototypical GRB SN Ic-BL SN 1998bw \citep[second panel;][]{Patat2001}, the prototypical LFBOT AT2018cow \citep[third panel;][]{Prentice2018}, the LFBOT AT2018gep \citep[fourth panel;][]{Ho2019}, the long GRB~071010A \citep[fifth panel;][]{Covino2008}, and the Einstein Probe transient EP~240414a \citep[sixth panel;][]{van_Dalen2024}. All the transients were converted to the redshift of EP~241021a (i.e., $z=0.7485$).}
    \label{fig:lc_comparison}
\end{figure}

\begin{figure}
    \centering
    \includegraphics[scale=0.53]{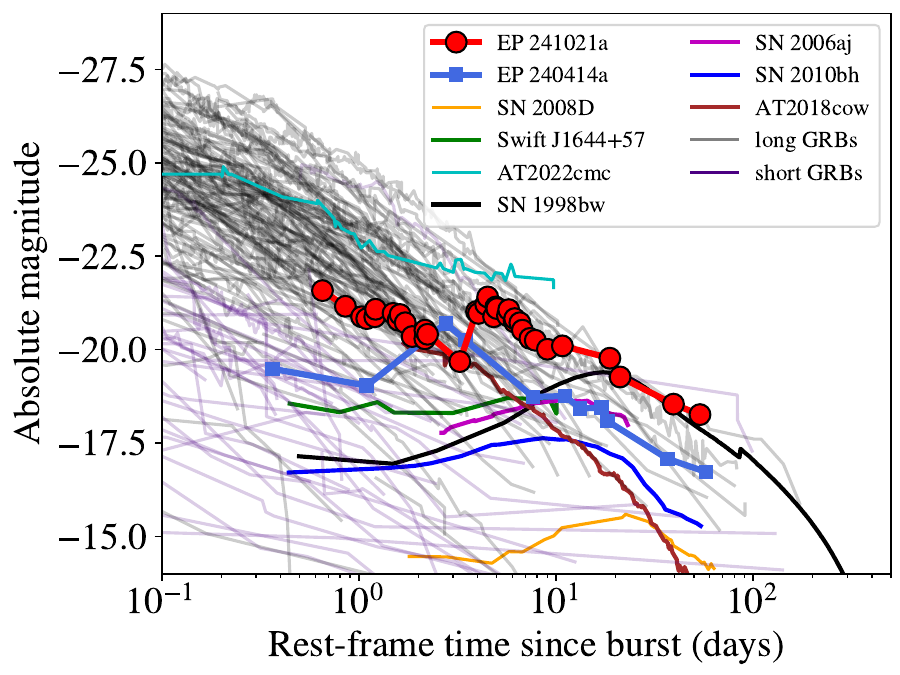}
    \vspace{-0.1cm}
    \caption{
    $r$-band optical light curves of FXT EP~241021a (red circles) compared to long and short GRBs \citep{Kann2006,Kann2010,Kann2011,Nicuesa2012,Kann2024}, the SN SBO event XRF 080109/SN 2008D, the X-ray flash source XRF 100316D/SN 2010BH, the LFBOT AT2018cow \citep{Xiang2021}, the jetted TDEs AT2022cmc and \emph{Swift}~J1644+57 \citep{Levan2011,Andreoni2022}, and the collapsar FXT EP~240414a \citep{van_Dalen2024}.}
    \label{fig:kann_plot}
\end{figure}

\section{Discussion}\label{sec:pheno}


\subsection{Light curve comparison}\label{sec:fen_LC}

Figure~\ref{fig:lc_comparison} shows a comparison between the light curve of EP~241021a and several well-known transients, such as GRB~060218/SN~2006aj \citep{Soderberg2006}, the GRB--SN Ic-BL SN~1998bw \citep{Patat2001}, the LFBOT AT2018cow \citep{Prentice2018} and candidate LFBOT AT2018gep \citep{Ho2019}, and the EP transient EP~240414a \citep{van_Dalen2024}, all shifted to a redshift of $z = 0.7485$.

The double peak of EP~241021a is reminiscent of that in the light curve of GRB~060218/SN~2006aj (see Fig.~\ref{fig:lc_comparison}, first panel from the top); however, neither the time of the peaks nor the luminosities do not match those observed in EP~241021a. Also, the early decay looks consistent with GRB afterglow (see Fig.~\ref{fig:lc_comparison}, fifth panel from the top, a comparison with GRB~071010A as an example). Indeed, the power-law index of $\alpha_1=-0.7\pm0.1$ during the first $\sim$6~days, is consistent with the optical afterglow decay of GRBs \citep[e.g.,][]{Melandri2008}.


During Epoch~II, EP~241021a exhibits a rapid re-brightening, peaking around day~$\sim 7.7$, followed by a power-law decay. From a GRB perspective, re-brightening within the first day is not uncommon in GRB afterglows \citep[e.g.,][]{Melandri2008,Martin-Carrillo2014,deUgarte2018}; however, the combination of brightness and timescale observed in EP~241021a is atypical for GRBs, which commonly evolve on timescales of thousands of seconds. Furthermore, although the prototypical LFBOT AT2018cow and EP~241021a differ in their timescales, the decay rate of EP~241021a after the main peak (see Fig.~\ref{fig:lc_comparison}, third panel from the top) is very similar to that of AT2018cow \citep[$\sim t^{-2.5}$;][]{Margutti2019}.
This behavior is similar to that observed in LFBOT and to the behavior of EP~240414a \citep[see Fig.~\ref{fig:lc_comparison}, bottom panel;][]{van_Dalen2024}. However, the colors of EP~241021a are not as blue as the colors observed for LFBOTs \citep[i.e., $ g-r<-0.2$~mag; see Fig.~\ref{fig:colors};][]{Ho2023}.

A mild re-brightening occurred around day~$\sim20$ in the $r$-band \citep[also visible perhaps in the $z$-band in][]{Busmann2024GCN}, which may result from a combination of multiple components. EP~241021a reached an absolute peak magnitude of $M_r \approx -20.1$. Although the high absolute magnitude at this peak is not consistent with that of AT2018gep \citep[which is a target that bridges LFBOTs and SNe~Ic-BL; see Fig.~\ref{fig:lc_comparison}, fourth panel from the top;][]{Pritchard2021}, the prototypical SN Ic BL SN~1998bw \citep[with a peak absolute magnitude of $M_R \approx -19.4$~mag;][]{Galama1998} matches the late-time evolution of EP~241021a reasonably well, especially beyond day $\sim20$ (see Figs.~\ref{fig:lc_comparison} second panel from the top).

Finally, Fig.~\ref{fig:kann_plot} shows a comparison of the $r$-band optical light curve of EP~241021a with various transients \citep[including GRBs, SNe, jetted TDEs, and LFBOTs;][]{Kann2006,Kann2010,Kann2011,Levan2011,Nicuesa2012,Ho2019,Xiang2021,van_Dalen2024}. Although the emission $\lesssim 1$~day was missed, Epoch~I is brighter than EP~240414a, supernovae, and AT2018cow. Nevertheless, it lies in a region typically occupied by optically faint long GRBs. The main peak of EP~241021a reaches an absolute magnitude similar to brighter optical afterglows; meanwhile, at late times, it remains brighter than typical optical afterglows and eventually matches the decay of SN~1998bw (beyond day $\sim20$ rest-frame).

To summarize, comparing the light curve of EP~241021a with other well-known transients, we identified that different epochs of EP~241021a match with various events at different episodes; for instance, at early and late times, EP~241021a is more consistent with GRB afterglow and the decay of SN~1998bw, respectively.

\begin{figure}
    \centering
    \includegraphics[scale=0.52]{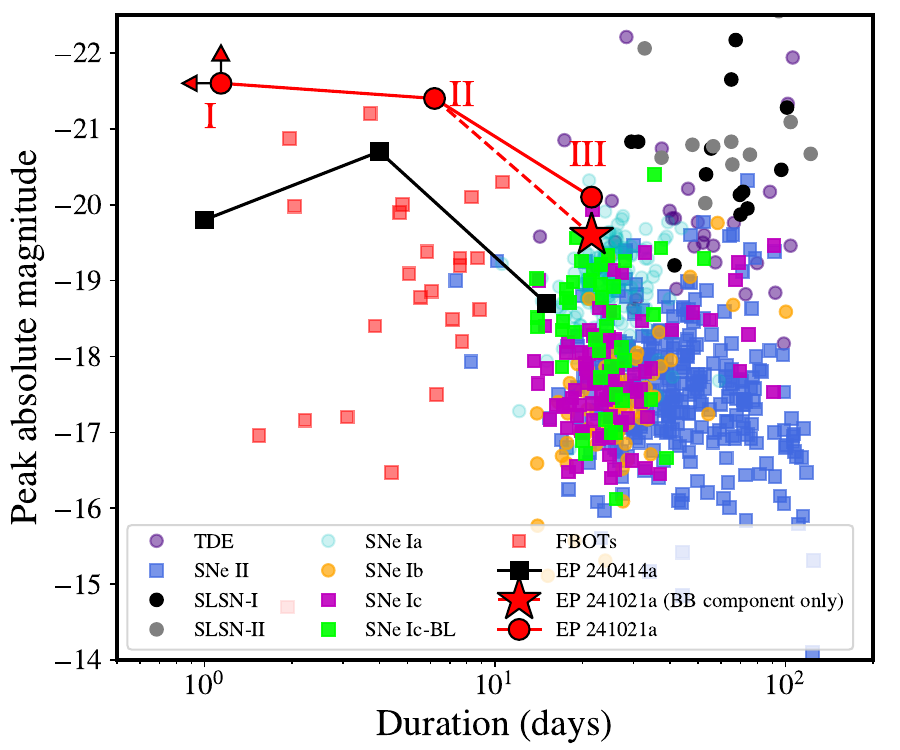}
    \caption{Absolute magnitude at peak versus the duration of transients above half-peak magnitude from the Zwicky Transient Facility Bright Transient Survey \citep{Perley2020} such as SN type Ia, Ib, Ic/BL, and tidal disruption events, and a set of fast transients \citep[AT2018cow-like sources;][]{Ho2023}. The Roman numbers depict the peak emission at each phase of EP~241021a. The red star represents the absolute magnitude considering only the thermal component (see \S\ref{sec:fen_SED}). Arrows in Epoch~I denote limits (see \S\ref{sec:fen_M_D}).}
    \label{fig:dura_peak}
\end{figure}

\subsection{Brightness-duration parameter space}\label{sec:fen_M_D}

The duration-absolute magnitude diagram is a powerful tool to distinguish between different classes of explosive events. Indeed, different classes of transients populate distinct regions of this parameter space: supernovae evolve slowly and have moderate luminosities, GRB afterglows flare briefly but are extremely bright at peak, while TDEs and super-luminous supernovae (SLSNe) remain luminous for months \citep[e.g.,][fig.~1]{Ho2023b}.
We use this diagram to explore the properties of EP~241021a \citep[e.g.,][]{Perley2020,Ho2023b,Ho2023}, taking into account the evolution of both parameters during its different epochs.
We approximate the peak during Epoch~I as the earliest detection of the target done by the KAIT telescope \citep{Zheng2024GCNa}, i.e., an absolute magnitude of $M_R\lesssim-21.6$~mag with a duration of $\lesssim1.14$ days. We should keep in mind that the peak could be brighter and the duration shorter.
In the case of the main peak during Epoch~II at $\sim7.7$~days, it reaches a peak absolute magnitude of $M_r\approx-21.4$~mag with a half-flux duration of $\approx6.2$~days (using a broken power-law fitting of the $r$-band light curve). 
Finally, the last re-brightening in Epoch~III, starting at day $\approx$19, yields an estimated absolute magnitude of $M_r\approx-20.1$~mag and a duration of $\approx21$~days.
Figure~\ref{fig:dura_peak} shows a comparison of the behavior of EP~241021a (roman numbers label its different epochs) and other transients (e.g., SNe Ia, Ib, Ic/BL, FBOTs, TDEs, and EP~240414a) in the duration--absolute magnitude parameter space. Overall, EP~241021a has different durations and absolute magnitudes during its different phases. The different episodes place this target in multiple regions in the parameter space. Epoch~I might even reside in a region populated by the GRB's afterglow, extrapolating to minutes to hours duration \citep[e.g., see][and its Figure~1]{Ho2023b}.
Moreover, the Epoch~II falls in a region just covered by FBOTs \citep{Ho2023}; however, it is brighter than the majority of the FBOTs reported by \citet{Ho2023}. This could be a consequence of a mixture of spectral components in the emission of the transient. 

Furthermore, the re-brightening of EP~240414a associated with a SN type Ic-BL \citep{van_Dalen2024} is fainter ($M_r\approx-18.5$~mag) than the peak at Epoch~III of EP~241021a. The peak of Epoch~III falls in a region mainly populated by TDEs, SN type Ia, and SLSNe. Only a few SNe Ic-BL from the sample \citep[e.g., SN~2021bmf and SN~2023wtq have a peak absolute magnitude of $M_r=-20.6$ and $-20.5$, respectively;][]{Perley2020,Aamer2023TNS,Anand2024} are brighter than EP~241021a. Unfortunately, in the case of EP~241021a, it was not possible to identify spectroscopic SN features. One possible explanation could be that the added contribution of several non-thermal spectral components masked the supernova spectroscopic features (see \S\ref{sec:modeling}).
To explore if Epoch~III of EP~241021a is consistent with SN emission, Fig.~\ref{fig:colors} shows a color comparison with four transients: SN~1998bw, AT2018cow, SN~2006aj, and EP~240414a. The colors at later epochs look more consistent with SN~1998bw and SN~2006aj than the others. A thermal (SN) contribution might be in line with the steep optical spectral index at day $\sim$37.

\begin{figure}
    \centering
    \includegraphics[scale=0.65]{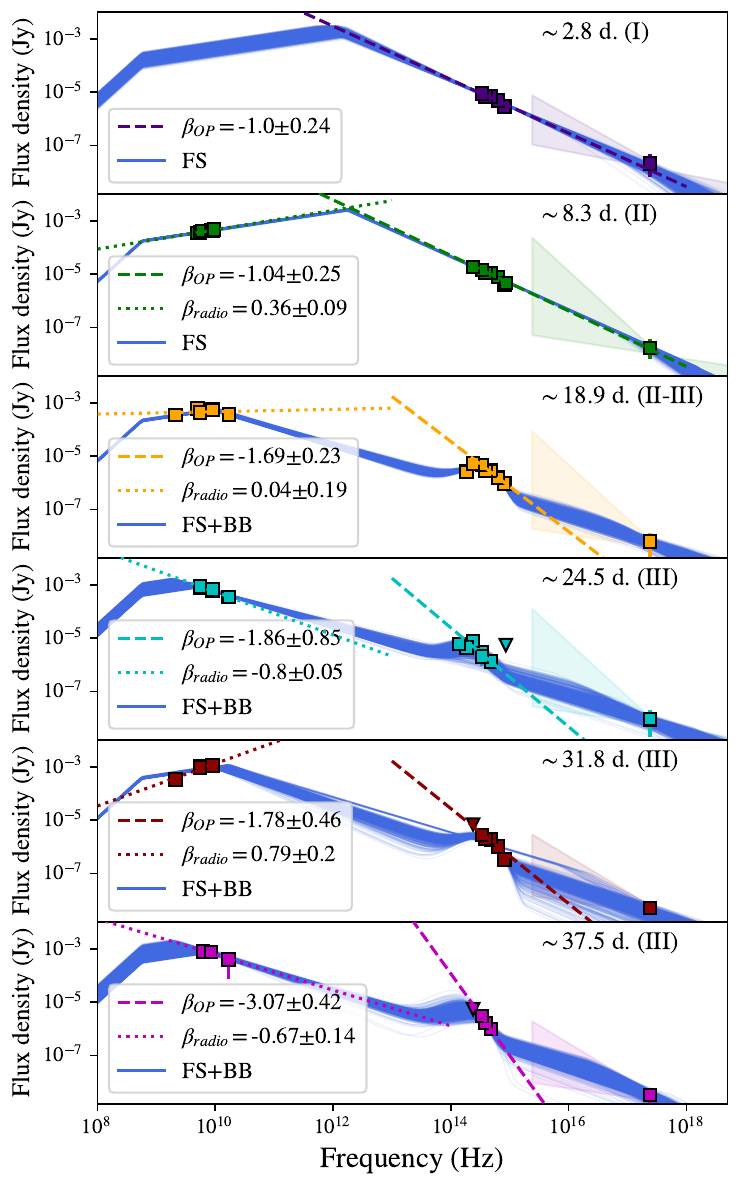}
    \vspace{-0.2cm}
    \caption{Radio-to-X-ray SED of the transient EP~241021a at six different stages. We fit a power-law function to the radio (dotted lines) and optical (dashed lines) data to visualize their trend. Moreover, we fit a more complex model (blue lines) using an MCMC method. In the first and second panels, we fit a forward shock (FS) model in the slow cooling regime ($\nu_a<\nu_m<\nu_c$, where $\nu_a$, $\nu_m$, and $\nu_c$ are the self-absorption, cooling, and characteristic synchrotron frequencies, respectively). In contrast, for the other panels a FS + black body (FS+BB) model is fit. The uncertainty on the X-ray spectral index is shown as a shaded region. We use the radio data from \citet{Gianfagna2025,Yadav2025}, optical from this work, NIR observations from \citet{Busmann2025,Gianfagna2025}, and the EP-FXT data during epoch $\sim32$~days from \citet{Shu2025}. Squares and triangle markers depict detections and upper limits, respectively. The optical and NIR data were corrected for Galactic extinction and for the contribution of the host galaxy.}
    \label{fig:SED_radio_xray}
\end{figure}

\subsection{SED analysis}\label{sec:fen_SED}

Based on the power-law light curves and optical spectra, and the optical-to-X-ray SED evolution throughout Epochs~I and II, it seems likely that both arise from a non-thermal afterglow spectral component. To explore this possibility, firstly, we compare the power-law indices from the optical light curves to those from canonical forward shock models \citep[FS;][]{Sari1999b,Granot2002b,Zhang2006,Gao2013b}.
Because the first optical observations of the transient were taken $\sim1.14$~day after the trigger, we assume that under the FS model in a deceleration relativistic regime before the jet break, the transient is in the slow cooling regime.
During Epoch~I, the decay rate 
points to an electron energy power-law index $p\approx2.0-2.3$ or $1.4-1.7$ in the regime $\nu_a<\nu_m<\nu<\nu_c$ (where $\nu_a$, $\nu_c$, and $\nu_m$ are the self-absorption, cooling, and characteristic synchrotron frequencies, respectively) in a constant interstellar medium (ISM) and wind density profile, respectively. Thus, we discard the wind profile due to the electron density of $p<2$. Taking the derived electron energy index and using the closure relations in the range $\nu_a<\nu_m<\nu<\nu_c$ \citep{Gao2013b}, $\beta\sim-0.7$, which is slightly similar to the X-ray-to-optical index derived from the SED (see Fig.~\ref{fig:SED}), supposing that X-ray frequencies are below $\nu_c$. Thus, we can conclude that a non-thermal FS emission is a consistent explanation at this stage.
Assuming this non-thermal emission originates from a relativistic jet characterized by an initial Lorentz factor ($\Gamma_0$), which decelerates through interaction with the ISM, we estimate $\Gamma_0$ based on the premise that the earliest optical detection marks an upper limit on the jet's deceleration time \citep[following the formalism of][]{Ghirlanda2018}. For an isotropic-equivalent kinetic energy of $E_0 = 10^{52}$~erg and a uniform ISM density with normalization parameter of $n_0 = 10$ and $1$~cm$^{-3}$ (see \S\ref{sec:modeling}), we derive a lower limit on the initial Lorentz factor of $\Gamma_0\gtrsim7-9$.

Furthermore, during the decay of the main peak of Epoch~II, we find a steeper slope, i.e., $\propto t^{-1.7\pm0.2}$ (excluding the data beyond day $\sim20$), where a post-jet-break scenario can be invoked to explain the steeper decay. In the post-jet break regime and in the frequency range of $\nu_a<\nu_m<\nu<\nu_c$ regime, the decay rate is consistent with $\alpha=3p/4$ \citep{Gao2013b}, and the electron density is $p\approx2.0-2.5$ considering an ISM; while in a wind medium $p<2$, discarding this profile again. 
On the other hand, the fast rise in the optical light curve from $\sim$5.6 to 7.7~days ($\propto t^{+3.6}$) cannot be explained based on the FS model, suggesting a more complex mechanism to be at work.
Furthermore, from the optical and X-ray SED evolution (see Fig.~\ref{fig:SED}), it is evident that both the optical and the X-ray emission can come from the same spectral component, at least until day $\sim18$; nevertheless, at later-times, although the X-ray data remains bright until day $\sim40$ \citep{Shu2025}, their physical origin might be different due to the mismatch of both spectral slopes (see Fig.~\ref{fig:SED}).

To verify a connection with an FS model and data from radio to X-ray wavelengths, we build the SED of the transient at different times using our and public multi-wavelength data from the literature: radio \citep[$\gtrsim2$~GHz;][see Table~\ref{tab:radio}]{Gianfagna2025,Yadav2025}, optical and NIR \citep[this work and][see Tables~\ref{tab:photometry} and \ref{tab:photometry_gcn}]{Gianfagna2025,Busmann2025}, and X-ray \citep[EP-FXT;][see Table~\ref{tab:x-rays}]{Shu2025} data. 
Figure~\ref{fig:SED_radio_xray} depicts the SED at six stages, covering Epochs~I, II, and III, where the dashed and dotted lines represent the best power-law fit of the optical and radio data, respectively, described by spectral indices $\beta_{\rm op}$ and $\beta_{\rm radio}$.

During days $\sim2.8$ and $8.3$ (see Fig.~\ref{fig:SED_radio_xray}, first and second panels from the top, respectively), it is clear that extrapolating the $\beta_{\rm op}$ relation to X-ray frequencies is consistent with the observed X-ray flux.
However, at day $\sim8.3$, extrapolating to lower radio frequencies, it vastly overpredicts the radio emission. Clearly, a break in the power-law extrapolation is needed. Meanwhile, extrapolating the radio power-law SED index ($\beta_{\rm radio}\approx+0.36$) to higher frequencies (optical) also overpredicts the optical flux. This behavior suggests that the optical and X-ray data can originate from the same spectral component, but it is not obvious that this is also the case for the radio data.
We explore whether a single FS model can explain the SED from radio to X-rays. To investigate this possibility, we implement a Markov Chain Monte Carlo (MCMC) method to fit the radio-to-X-ray data of six stages (covering Epochs~I, II, and III; see Fig.~\ref{fig:SED_radio_xray}) using a relativistic, isotropic, self-similar decelerating shock model under the slow-cooling regime. In this regime, assuming the convention $F_\nu \propto \nu^\beta$, and defined by the frequencies $\nu_a$, $\nu_m$ and $\nu_c$ and the maximum peak flux $F_{\rm max}$, the spectral slopes follow the standard synchrotron slopes as: $\beta = +2$, $+1/3$, $-(p-1)/2$, and $-p/2$. Due to the lack of low-frequency radio observations, we fix the self-absorption frequency at $\nu_a\approx10^9$~Hz \citep[see][]{Zhang2006}.
As shown by the blue solid lines (posteriors) in the top panel of Fig.~\ref{fig:SED_radio_xray}, the FS model successfully reproduces the SED from optical to X-rays of the transient at day $\sim2.8$ (i.e., Epoch~I). The best-fit parameters are: $p = 3.0\pm0.2$, $\log(\nu_m) = 12.4^{+0.1}_{-0.3}$, $\log(\nu_c)=18.2_{-1.6}^{+0.8}$, and peak flux $F_{\rm max} = 2.5_{-1.1}^{+0.7}$~mJy. Moreover, during day $\sim8.3$ (see Fig.~\ref{fig:SED_radio_xray}, second panel from the top), the FS model also matches the broadband SED of the transient, from radio to X-rays. The best-fit parameters are slightly different with respect to those at the previous stage: $p = 3.0\pm0.1$, $\log(\nu_m) = 12.5^{+0.0}_{-0.02}$, $\log(\nu_c)=17.9_{-1.2}^{+1.1}$, and $F_{\rm max} = 2.6\pm0.2$~mJy. Both fits show that afterglow-like emission could explain the entire SED of the transient during Epochs~I and II, strengthening the non-thermal nature of the counterpart at both epochs.


In the subsequent epochs, we discover (see Fig.~\ref{fig:SED_radio_xray}) that $\beta_{\rm op}$ becomes steeper, and $\beta_{\rm radio}$ shifts to positive values (due to the lack of observation $\gtrsim9$~GHz on day $\sim31.8$, this transition is not fully resolved). As we demonstrated above, the evolution to a steep spectral index in the optical part of the SED beyond day $\sim20$ is incompatible with non-thermal emission (see Fig.~\ref{fig:SED}). If we try to explain the evolution in the SED using a single FS model, we do not obtain a good fit.
Furthermore, the slight re-brightening in the optical at $\gtrsim18$ days can be explained by the emergence of an additional spectral component. We interpret this new component as blackbody (BB) emission, which can explain the steeper optical spectral index observed at late times (in agreement with the Wien tail of the BB spectrum), and it also clarifies why the X-ray spectral index differs from that of the optical (see Fig.~\ref{fig:SED}, last three panels from the top).
To test this hypothesis, we fit a simple FS model combined with a BB component, described by a temperature and radius $T_{\rm BB}$ and $R_{\rm BB}$, respectively. This combined model is labeled as FS+BB. We expect a transition around days $\sim18-19$, when the non-thermal component still plays a significant role in the SED. Using the data obtained on day $\sim18.9$, we perform fits using both the simple FS model and the FS+BB model, and evaluate their performance using the \emph{Bayesian Information Criterion} (BIC)\footnote{BIC $= -2\ln{\mathcal{L}} + k\ln{N}$, where $\mathcal{L}$ is the maximum likelihood of the data, $k$ is the number of model parameters, and $N$ is the number of data points \citep{Ivezic2014}. We consider the threshold criterion of \hbox{$\Delta$BIC${=}$BIC$_{h}$-BIC$_{l}{>}2$} to discriminate when comparing two different models, where BIC$_{\rm h}$ and BIC$_{\rm l}$ are the higher and lower model BIC, respectively. The larger $\Delta$BIC, the stronger the evidence against the model with a higher BIC is \citep{Liddle2007}.}.
Based on the BIC values, we find that the FS+BB model provides a significantly better fit (see Fig.~\ref{fig:SED_radio_xray}, third panel from the top) to the multi-wavelength data during day $\sim18.9$ (BIC$_{\rm FS} = -134.99$, BIC$_{\rm FS+BB} = -284.82$).

Extending our analysis to the epochs from approximately $\sim24.5$ to $37.5$~days (the posteriors are in Table~\ref{tab:posteriors}), we find that the FS+BB model can simultaneously explain the spectral evolution observed in the radio, optical, and X-ray bands (see Fig.~\ref{fig:SED_radio_xray}). During this observations on Epoch~III ($\Delta 
t\approx23$~days; see Fig.~\ref{fig:SED_radio_xray}, the last three panels from the top), the blackbody (BB) temperature cools from $\sim 9000$ to $5300$~K, while its bolometric luminosity remains nearly constant at $L_{\rm bol} \sim(1.0 - 1.4)\times 10^{44}$ erg~s$^{-1}$. 
Regarding the FS model parameters, we find that $F_{\rm max}$ and $\nu_c$ evolve consistently with expectations for a post-jet-break afterglow, i.e., $F_{\rm max} \propto t^{\sim-1}$ and $\nu_c \propto t^{\sim0}$ \citep{Sari1999b}, at 95\% confidence level. However, the frequency $\nu_m$ decays faster than expected: $\nu_m \propto t^{\sim-2.6}$, compared to the standard jet-break model, i.e., $\nu_m \propto t^{-2}$ \citep{Sari1999b}. This discrepancy could be due to the emergence of additional non-thermal components (see Section~\ref{sec:modeling}), indicating to a more complex structure than a unique non-thermal component \citep[which we will also address in \S\ref{sec:modeling}, e.g.,][]{Filgas2011,Resmi2012}, evolving microphysical parameters \citep[e.g.,][]{Greiner2013,Varela2016,Marongiu2022}, or variations in the circumburst medium such as density jumps or cavities \citep[e.g.,][]{Harrison2001,Lazzati2002,Huang2007,Gat2013}.
Indeed, \citet{Gianfagna2025} reported the presence of a second emission component in the radio band (based on ALMA data at $\sim10^{11}$~Hz) for EP~241021a, first detected around day $\sim50-70$. 

\begin{figure}
    \centering
    \includegraphics[scale=0.65]{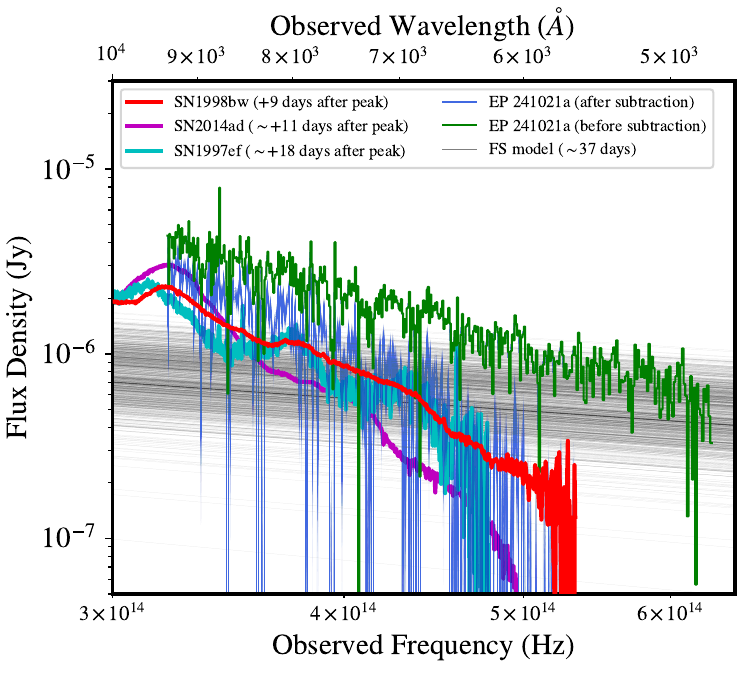}
    \vspace{-0.2cm}
    \caption{MUSE spectrum of FXT EP241021a before (green) and after (blue) subtracting our best-fit afterglow model. We show spectra of SN~1998bw \citep[red line;][]{Patat2001} to show that it matches not only the shape but also the overall flux level of our spectrum, and the best-fit SED at this epoch (see \S\ref{sec:fen_SED}). Moreover, as a comparison of the shape and continuum of the transient, we scaled and overplotted the SN~2014ad and SN~1997ef \citep{Sahu2018,Modjaz2014}. For comparison, we overplot a random sample of posteriors from the FS model (black lines; see \S\ref{sec:fen_SED}).}
    \label{fig:type_Ic}
\end{figure}

\subsection{EP241021a as a tidal disruption event}

Based on the high peak X-ray emission of the transient detected by WXT-EP (i.e., $L_{\rm X,peak}\sim10^{48}$~erg/s) and its duration \citep[$\sim100$~sec.;][]{Shu2025}, it is possible that the transient is related to a jetted TDE, which could explain the early non-thermal emission present in the transient. If we compare the X-ray emission of EP~241021a and the jetted TDE Sw~J1644+57 and AT2022cmc \citep[see Fig.~\ref{fig:xray_plot};][]{Bloom2011,Levan2011,Andreoni2022}. The initial WXT-EP detection is consistent with the early luminosity of Sw~J1644+57; nonetheless, after $\sim1$~day of the trigger (rest-frame), the X-ray emission is fainter than Sw~J1644+57 by two orders of magnitude. Furthermore, the light curve of EP~241021a does not show evidence for the strong variability detected in Sw~J1644+57. The jetted TDE AT2022cmc \citep{Andreoni2022} is brighter than EP~241021a, and follows a steeper X-ray decay \citep[$\propto t^{-2.1}$;][]{Eftekhari2024} than EP~241021a. At optical wavelength, AT2022cmc decays steadily while it remains brighter than EP~241021a (see Fig.~\ref{fig:kann_plot}); meanwhile, in the optical range, Sw~J16644+57 remains fainter than EP~241021a (see Fig.~\ref{fig:kann_plot}). Although EP~241021a does not follow the same evolution as Sw~J1644+57 and AT2022cmc, EP~241021a might be fully consistent with the large spread in behavior shown by relativistic TDEs \citep[e.g.,][]{Teboul2023,Yuan2025b}.

Black holes (BH) with masses more than $\sim10^5$~$M_\odot$ swallow typical WDs whole, and only those of lower mass can produce tidal disruptions of WDs. Many of WD-IMBH TDE systems are also predicted to launch relativistic jets \citep{Strubbe2009,Zauderer2011,DeColle2012} as a result of the extremely rapid mass supply \citep[e.g.,][]{DeColle2012,Krolik2012,Shcherbakov2013}, powered by rapid IMBH spin (which has been estimated in some IMBH candidates, e.g.,\citealp{Wen2021} and \citealp{Cao2023}) and high magnetic flux \citep[e.g., from a magnetic WD;][]{Cenko2012, Brown2015,Sadowski2016} via the \citet{Blandford1977} mechanism. 
\citet{MacLeod2016} shows typical peak jet luminosities between $\sim10^{47}-10^{50}$~erg~s$^{-1}$ and rise timescales of $\sim10^2-10^4$~s, in agreement with the initial X-ray duration of EP~241021a. Under this assumption, the X-ray emission at $\gtrsim1$~day (rest-frame) of $5\times10^{44}$~erg~s$^{-1}$ would be related to a thermal accretion disk emission; however, this value is $\sim1-2$ orders of magnitude higher than the expected Eddington luminosity of a BH with mass $\sim10^4$~$M_\odot$ \citep{Frank2002}.

Several studies predict a wide diversity of electromagnetic signatures depending on the orbital parameters of the WD--IMBH system. For instance, single, strongly disruptive passages are thought to produce quick-peaking light curves with power-law decay tails as debris slowly falls back to the IMBH \citep[e.g.,][]{Rosswog2009,Shcherbakov2013}. It has also been suggested that such encounters may result in the WD detonation \citep{Rosswog2009,Haas2012,MacLeod2014}, and such a detonations imply that besides emission due to the accretion disc, where a Type Ia-like SN peaking at optical wavelength could be present \citep[e.g.,][]{Rosswog2008a,Rosswog2008b,MacLeod2016}.
Figure~\ref{fig:IMBH_TDE} compares the light curve of EP~241021a and those predicted from the hydrodynamical simulations of \citet{MacLeod2016}, who consider a wide range of viewing angles. The optical counterpart of EP~241021a is brighter than these predictions, including at later epochs when the thermal component is arising. In addition, the color evolution of the transient deviates from the predictions of \citet{MacLeod2016}. Thus, we conclude that EP~241021a is not likely to originate from a WD-IMBH TDE where the WD detonated.

\subsection{Collapsar origin}

The optical colors of the late BB-like emission of EP~241021a are similar to those of collapsar events such as SN~1998bw \citep{Clocchiatti2011} and SN~2006aj \citep{Campana2006}, suggesting that it might be due to a supernova (see Fig.~\ref{fig:colors}). Furthermore, its optical light curve is consistent with the evolution of SN~1998bw beyond day $\sim20$ (see Fig.~\ref{fig:kann_plot}). To explore the detectability of a potential supernova component, we estimated the brightness of a redshifted SN~1998bw analogue (at $z = 0.7485$ and $K$-corrected). At $\sim$25~days after the trigger ($\approx$15 days rest-frame), such an event would reach $r\approx23.6$~mag, whereas EP~241021a was observed at $r\approx22.9$~mag. Therefore, the transient was about twice as luminous as an SN 1998bw-like explosion at the same phase, indicating the presence of an additional component that might dilute the expected broad-lined Type Ic SN absorption lines.

We estimate the contribution of the thermal component to the total emission\footnote{To estimate the contribution of the BB emission to the total emission, we compute the ratio $\int BB~d\nu/\int(FS+BB)d\nu$, in the frequency range of the $r$-band.} at $\sim18.9$ and $37.5$~days (i.e., $\sim10.4$ and $21$~days rest-frame) to be $\approx50$ and $85$\%, respectively, at $r$-band frequencies.
Therefore, the peak absolute magnitude of the thermal component during Epoch~III in the $r$ band is $M_r\approx-19.6$~mag based on a contribution of $\approx65$\% during Epoch~III. Comparing this value with the absolute magnitude of Type Ic SNe (e.g., see Fig.~\ref{fig:dura_peak}, where the red star depicts the peak absolute magnitude of the thermal contribution), we find that the thermal component by itself is consistent with a Type Ic-BL SN. 

Finally, we investigate the MUSE spectrum in detail because during this stage, based on our work, the thermal component contributed $\sim85$\% to the total emission, which is higher than the factions determined for the preceding observations.
We performed a joint fit of a fixed SN template (SN~1998bw spectrum at day $+9$ after the peak) scaled to the redshift of EP~241021a in combination with an afterglow emission modeled as a power-law model to mimic a SN + non-thermal contribution to the transient. Also, we assume Gaussian noise to emulate the noise level in the MUSE spectrum of EP~241021a \citep[corrected by Galactic extinction using the dust model of][]{Calzetti2000}. 
Figure~\ref{fig:type_Ic} presents the MUSE spectrum of EP~241021a before (green line) and after (blue line) subtracting the best-fit afterglow component. The best-fit afterglow has a spectral decay (i.e., $F_\nu\propto\nu^{-0.6}$) consistent with that derived from fitting the SED of the transient (see Fig.~\ref{fig:type_Ic}). We compare our afterglow-subtracted spectrum with the SN~1998bw spectrum \citep[scaled only by the distance;][]{Patat2001}, SN~2014ad \citep{Sahu2018}, and SN~1997ef \citep[][not only scaled by the distance]{Modjaz2014}. Overall, although the S/N of the MUSE spectrum does not allow for identifying SN absorption features, the afterglow-subtracted spectrum is consistent with the spectral shape and continuum of SN~1998bw and other SNe Ic-BL, indicating that the late-time optical counterpart of EP~241021a might well be consistent with this type of collapsar transients.


\begin{figure}
    \centering
    \includegraphics[scale=0.65]{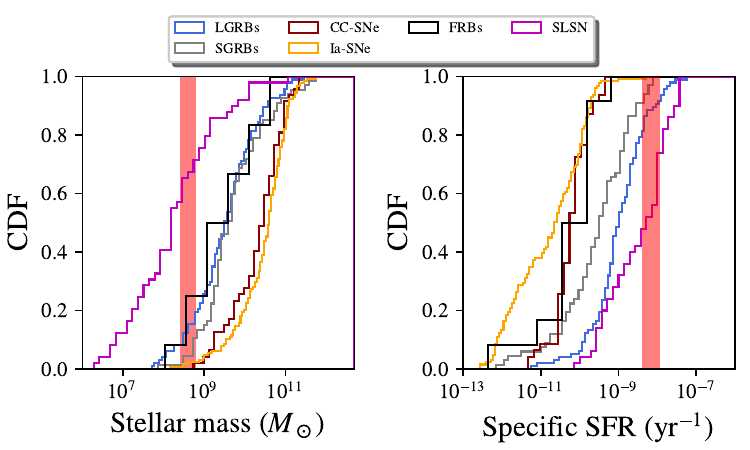}
    \vspace{-0.2cm}
    \caption{Comparison of the host-galaxy properties of EP~241021a at 68\% confidence level (red region) with the cumulative distributions of 
    stellar mass (left panel) and 
    specific star-formation rate (right panel) 
    for LGRBs \citep{Li2016, Blanchard2016}, 
    SGRBs \citep{Fong2010, Fong2012, Fong2013, Margutti2012b, Sakamoto2013, Berger2013b, Fong2022, Nugent2022}, 
    FRBs \citep{Heintz2020}, 
    CC-SNe and Ia-SNe \citep{Tsvetkov1993,Prieto2008,Galbany2014}, and 
    SLSNe \citep{Schulze2021}.
    }
    \label{fig:host_comparison}
\end{figure}

\subsection{Host galaxy properties} \label{sec:host}

To detect the host galaxy of EP~241021a, we used HiPERCAM data taken $\sim$304 days after the X-ray trigger. At that epoch, the light of the transient had vanished. The host galaxy of the transient is extended, and it has a best-fit position consistent with
RA$_{\rm J2000.0}=$01$^{\rm h}$55$^{\rm m}$23$^{\rm s}$.44,  
Dec$_{\rm J2000.0}=$+05$^\circ$56$\arcmin$17$\arcsec$.91. We measure the following photometric points: $m_u=25.40\pm0.23$, $m_g=24.98\pm0.09$, $m_r=24.38\pm0.09$, $m_i=24.48\pm0.17$, and $m_z=23.43\pm0.33$~AB~mag. Therefore, the transient's photometry becomes similar to the host galaxy beyond day $\sim70$.

Moreover, the offset between the transient and the centre of the host galaxy is $0.45\arcsec\pm0.67$\arcsec (which means a physical offset of $3.39\pm5.06$~kpc) and a chance alignment probability of $0.004$.
We derive the host properties of EP~241021a through the SED fitting of the combined photometry, considering a star-formation history (SFH) model described by a delayed exponentially declining function using the package \texttt{Bagpipes} \citep{Carnall2018,Carnall2019}.
Figure~\ref{fig:host_sed} depicts the best-fitting SED of the available photometric data using \texttt{Bagpipes}. We obtain a stellar mass, star-formation rate (SFR), metallicity, mass-weighted age, and dust attenuation of $\rm \log(M_*/M_\odot)=8.6\pm0.2$, $\rm SFR=2.8_{-1.1}^{+1.5}$~$M_\odot$~yr$^{-1}$, $\rm \log(Z/Z_\odot)=-0.1\pm0.3$ \citep[or 12+$\rm \log(O/H)=8.6\pm0.3$, using][]{Asplund2009}, $t_m=0.07_{-0.03}^{+0.14}$~Gyr, and $\rm A_V=1.1\pm0.3$, respectively. 
These results suggest a low-mass, young galaxy with moderate star formation activity, whose metallicity is close to or slightly above solar.
Furthermore, it shows a moderate extinction, consistent with active star formation. The high metallicity of the host galaxy is uncommon for the typical long GRB host galaxies \citep[e.g.,][]{Levesque2014,Hashimoto2015}. The model photometry obtained by \texttt{Bagpipes} is
$m_u= 25.21$, $m_g= 25.01$, $m_r= 24.52$, $m_i= 24.23$, $m_z= 23.79$, $m_J= 23.76$, $m_H= 23.40$, and $m_{Ks}= 23.32$~AB~mag, used to correct the contribution of the host galaxy to photometry. Small band-to-band differences (e.g., in the $u$-band) are consistent with measurement uncertainties, while the generally brighter measured magnitudes suggest residual transient contribution, thereby justifying the use of the host model photometry for subtraction.

In Figure~\ref{fig:host_comparison}, we compared the stellar mass ($M_*$) and specific star formation rate (sSFR) of EP~241021a with other transients such as long and short GRBs, CC- and type Ia SNe, and SLSNe. The cumulative distributions of $M_*$ and sSFR for different transient host populations show that the host of EP~241021a lies at relatively low stellar mass, around  $\sim4\times10^{8}$~$M_\odot$, and exhibits a high sSFR of $\sim9\times10^{-9}$~yr$^{-1}$. 
This combination places the galaxy below the typical stellar masses of Type Ia SNe and FRB hosts, while aligning more closely with the properties of long GRB and SLSN hosts, which are preferentially found in actively star-forming galaxies from moderate to high star formation rates.
Indeed, long GRBs are preferentially associated with irregular, star-forming, and low-metallicity galaxies, with a few being spirals with active star-formation \citep{Fruchter2006,Stanek2006}. Their host galaxies are relatively metal-poor compared to the field population \citep{Fynbo2003,Prochaska2004,Fruchter2006}, favoring a collapsar progenitor model \citep{Woosley2006}. On the other hand, the formation efficiency of merger-driven transients should be independent of metallicity \citep{Giacobbo2018,Chruslinska2018,Neijssel2019}; indeed, the metallicities of short GRB hosts are significantly higher than those of long GRB hosts \citep{Berger2014,Modjaz2008,Levesque2010,Levesque2014}.
This indicates that the EP~241021a’s environment is characteristic of young, star-forming systems and points toward a massive-star progenitor channel, rather than one associated with older stellar populations, aligned with our previous results.

\subsection{Event rate density} \label{sec:rates}

In this section, based on the unique properties 
of EP~241021a, such as the lack of gamma-ray counterparts, light curve evolution, peak X-ray luminosity, and progenitor, we estimate its local event rate density. The estimated local event density ($\rho_0$) can be derived from \citep{Zhang2018,Sun2019}:
\begin{equation}
N_{\rm WXT}=\frac{V_{\rm max}\Omega_{\rm WXT} T_{\rm WXT}}{4\pi}\rho_0, 
\end{equation}
where $V_{\rm max}$ represents the maximum volume at which equivalent objects could be detected by EP, $\Omega_{\rm WXT}$ and $T_{\rm WXT}$ depict the EP-WXT FoV (i.e., 3600~square degrees) and on-sky time of the satellite (i.e., $\sim1.5$~yrs until mid of 2025), respectively, and $N_{\rm WXT}$ is the total number of sources detected (i.e., one event in this case).
EP-WXT can detect and identify X-ray transients as faint as $F_{\rm X}{\sim}10^{-10}$~erg~cm$^{-2}$~s$^{-1}$ for a typical exposure time of $\sim200$~seconds. Adopting a peak luminosity of $L_{\rm X}{\sim}10^{48}$~erg~s$^{-1}$ from 
EP~241021a, EP-WXT can then detect similar objects to redshifts up to $z_{\rm max}\sim1.3$. 
Moreover, following the methodology of \citet{Sun2024}, the maximum volume $V_{\rm max}$ is
\begin{equation}
    V_{\rm max}=\int_0^{z_{\rm max}}\frac{\Omega_{\rm WXT}}{4\pi}\frac{f(z)}{(1+z)}\frac{dV(z)}{dz}dz
\end{equation}
where, $f(z)$ is a weighted star-formation history function \citep[taken from][]{Madau2014}, while $dV(z)/dz$ is given by
\begin{equation}
    \frac{dV(z)}{dz}=\frac{c}{H_0}\frac{4\pi D_L(z)^2}{(1+z)^2\sqrt{\Omega_M(1+z)^3+\Omega_\Lambda}},
\end{equation}
where $c$, $H_0$, and $D_L$ are the light speed, Hubble constant, and luminosity distance, respectively, while $\Omega_M$ and $\Omega_\Lambda$ are matter and dark energy density of the Universe, respectively. In this way, and considering Possonian uncertainties, we obtained a local event rate density of $\rho_0\approx0.23_{-0.17}^{+0.34}$~Gpc$^{-3}$~yr$^{-1}$ at 68\% confidence level. 
Also, if we consider a beaming correction factor of $f_b^{-1}\sim30$ (where $f_b\simeq\theta_j^2/2$), which corresponds to a mean jet opening angle of $\theta_j\sim15^\circ$ (see \S\ref{sec:modeling} for more details), the local event rate density is $\rho_0\approx6.9_{-5.1}^{+10.2}$~Gpc$^{-3}$~yr$^{-1}$. This value is much lower than the rate of other potential extragalactic progenitors of FXTs, such as the merger rate of 
BNS systems \citep[$\rho_0^{\rm BNS}{=}320_{-240}^{+490}$~Gpc$^{-3}$~yr$^{-1}$;][]{Abbott2021a}, low-luminosity long GRBs \citep[$\rho_{\rm 0,LL-LGRBs}{\sim}$150--2100~Gpc$^{-3}$~yr$^{-1}$ (here $f_b^{-1}{\sim}1$--14 was used;][]{Liang2007,Zhang_book_2018}, 
long GRBs \citep[$\rho_{\rm 0,LGRBs}{=}$40--380~Gpc$^{-3}$~yr$^{-1}$, using $f_b^{-1}{\sim}50$--500;][]{Wanderman2010,Palmerio2021} and 
SGRBs \citep[$\rho_{\rm 0,SGRBs}{=}$45--200~Gpc$^{-3}$~yr$^{-1}$, using $f_b^{-1}{\sim}25$--110;][]{Guetta2005b,Wanderman2015}. 
Also, our outcome is lower than the FXT-SNe estimation from \citet{Rastinejad2025}.
However we should consider our rate density as a lower limit, given the fact that several other FXTs detected by EP have not been classified, and their origin is still unknown.
This lower limit is consistent with the millisecond magnetar formation rate from massive stars \citep{Biswas2025}. This is an interesting possibility, given that a magnetar engine in EP~241021a is an option to explain the rising flux in optical and X-ray bands at day $\sim$8 (see \S\ref{sec:modeling} for more details).

\begin{figure*}
    \centering
    \includegraphics[scale=0.52]{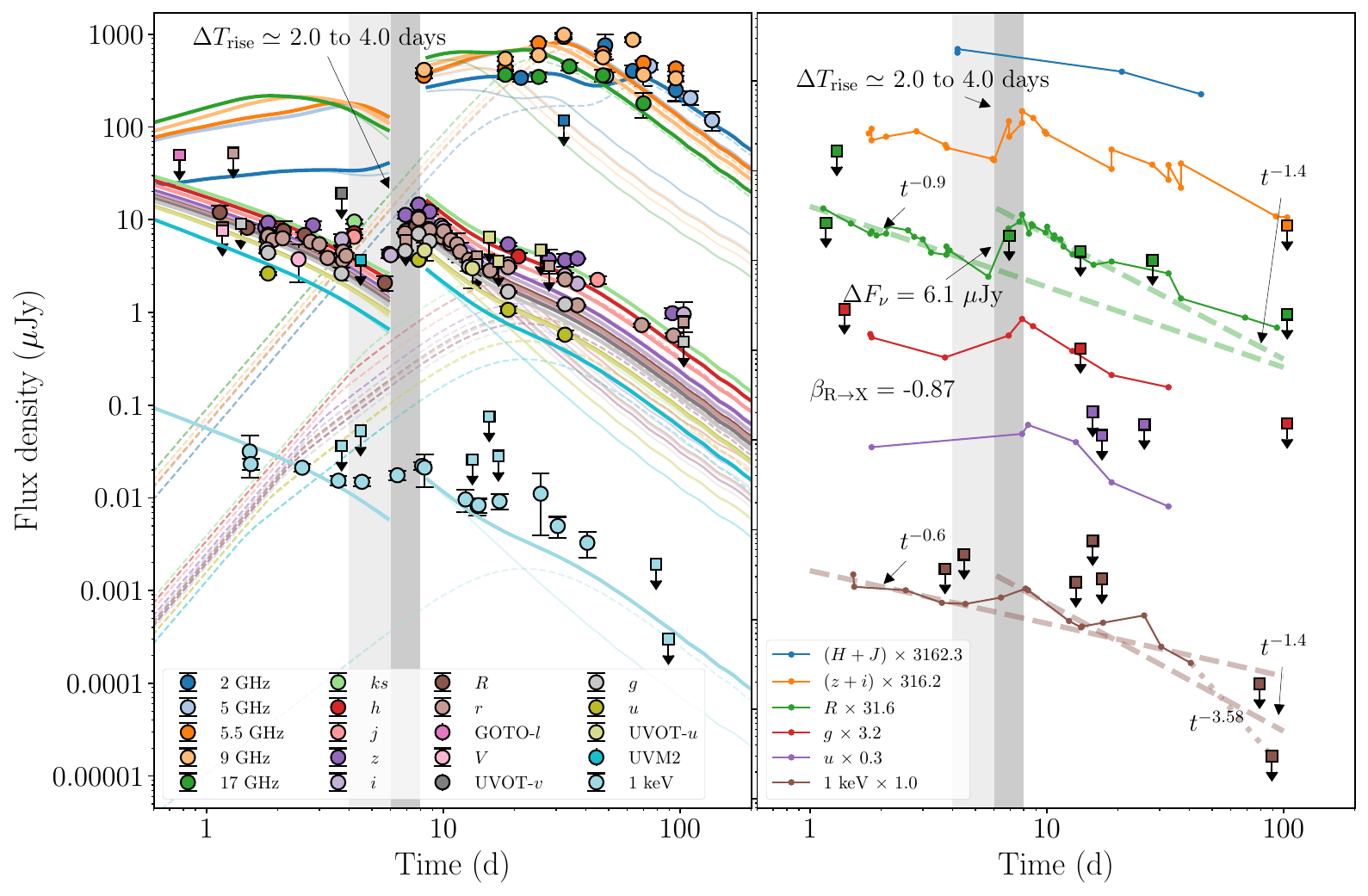}
    \vspace{-0.1cm}
    \caption{Multi-wavelength light curve modeling of the transient EP~241021a. For modeling the multi-wavelength light curves, a minimum of three non-thermal components is necessary: 
    $i)$ at $t\lesssim7$~days during Epoch~I (solid lines), a regular GRB afterglow (optical and X-ray data); 
    $ii)$ at $t\gtrsim8$~days during Epoch~II (narrow solid lines), a "regular" GRB afterglow with increased energy relative to the first component, possibly due to rapid energy injection between 6 and 8 days; 
    and $iii)$ during Epoch~III, a very slow component, that begins to contribute significantly to the emission at $t>10$~days (dashed lines), is responsible for most of the radio afterglow. This component has a Lorentz factor on the order of $\Gamma\approx3.5$.
    The dark grey vertical panel indicates the energy injection period for the rapid flux increase, in the $r$-band, while the light grey panel indicates the potential earliest start of the flux increase in other bands.}
    \label{fig:aft_model}
\end{figure*}

\section{Light curve modeling} \label{sec:modeling}

\subsection{Epochs I and II: non-thermal afterglow modeling}

To model the light curve, we group the various observations into frequency ranges for a simpler analysis, which we label: $H + J$ ($H$- and $J$-band); $z+i$ near infrared ($z$- and $i$-band); $R$ for red ($R$- and $r$-band); $g$ for visible ($g$-band); $u$ for ultraviolet ($u$-band); and 1~keV for X-ray at 1 keV.
These are plotted in the right panel of Figure~\ref{fig:aft_model}, with data points represented as dots without uncertainty and upper-limits shown as squares with arrows.
For the $R$ and 1~keV data, we fit two power-law models, one at $\lesssim6$~days (Epoch~I) and a second one at $\gtrsim8$~days (Epoch~II).
The power-law indices for these fits are noted within the figure:
the early, $\lesssim6$~days data at $R$ is $\propto t^{-0.9}$;
the early X-ray data is $\propto t^{-0.6}$;
the late, $\gtrsim8$~days data at $R$ and X-ray is $\propto t^{-1.4}$;
while the very late X-ray limit requires $\propto t^{<-3.58}$ after the last detection.

There are different decline indices between optical and X-ray at early times, where the X-ray shows a slower decline than the optical, which is unusual for GRB afterglows \citep{Zaninoni2013}.
However, if a blastwave decelerates within a wind-like medium, then the X-ray could show such a different temporal decline, where 1 keV $> \nu_c$.
For a decline index of $\alpha = -0.6$, the wind medium $\nu > \nu_c$ standard closure relation would require an accelerated electron power-law index of $p \sim 1.47$. Nonetheless, the standard closure relation is not valid for $p<2$, and the valid solution is unphysical for $\alpha = -0.6$ \citep{Gao2013}. We therefore rule out a wind medium as the origin of the mismatch between the decline index of the optical and X-ray light curves.
Figures~\ref{fig:SED} and \ref{fig:aft_model} indicate an optical to X-ray spectral index $\beta\sim -0.9$. If we apply this to the X-ray data, with 1 keV $> \nu_c$, then the value for $p = 1.8$ gives $\alpha = -0.96$, which is similar to that found using a simple power-law fit to the optical data.
For $\nu_m < \nu < \nu_c$, the decline index would be $\alpha = -0.71$. Considering the short timescale variability in the optical at Epoch~I, this value seems more reasonable, although not a precise match to the $t^{-0.9}$ power-law fit to the data in $R$.

The light curve discontinuity between $\sim$6 and 8 days results in a jump in flux, which could for instance be due to energy injection due to a refreshed shock, a jet, or a highly energetic jet core region (that is initially off-axis to the observer) becoming visible, or a patchy shell (where the energy varies randomly across the jet head resulting in `hot spots' or `sub-jets').
The duration of the rise is $\Delta T_{\rm rise} \simeq 2$~days and peaks at $\sim$8~days in the observer frame.
The short rise timescale for the flux increase at 6 -- 8 days violates the expectation for a refreshed shock $(\Delta T > T/4)$ or a patchy shell $(\Delta T > T)$ \citep{Ioka2005}.
An off-axis jet, a patchy shell/sub-jet, or a core region coming into view may be able to explain the rapid rise, however, given the sharp turn-around in the afterglow temporal structure at $>8$~days, the angular width of the off-axis jet would have to be extremely narrow (sharpness of the peak) and have a very sharp jet edge (the steepness of the rise).
Such conditions are not expected, even in a sub-jet model \citep{Nakamura2000}.
We therefore do not attempt to model here the details of the short rise timescale with a physical model, but leave it to future work.
However, we point out that at $\sim 4$~days post-burst the X-ray afterglow begins to deviate from the steady and continuous decline seen for the optical data in Epoch~I (the energy injection timescale for the afterglow is highlighted within Fig.~\ref{fig:aft_model}, grey regions).

The spectral index between optical and X-ray at $> 10$~days is $\beta_{\rm ox} \sim -1$, giving a $p = 2$ for Epoch~II.
The rapid decline between $\sim$10 -- 20 days would require a post-jet-break evolution for a $p=2$, $\beta_{\rm ox} = -1$ SED, yet the late X-ray excess (as well as the radio data) requires a longer-lived component.
This may be due to a second energy injection episode, or alternatively, the emission from a wider jet component coming into view.

\begin{figure*}
    \centering
    \includegraphics[scale=0.35]{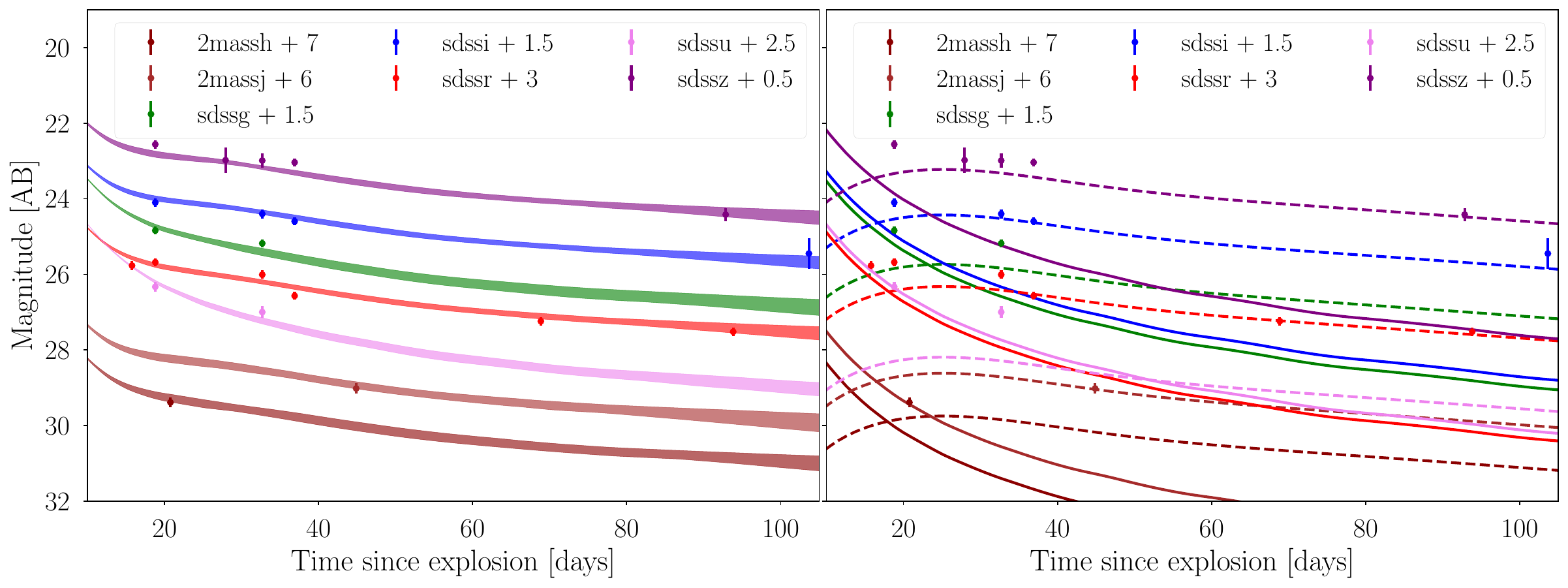}
    \vspace{-0.1cm}
    \caption{Multi-wavelength light curve modeling of the transient EP~241021a beyond day $\sim15$. In the left panel, we show $90\%$ credible intervals for different bands from our fit to the optical data with a thermal (one-zone radioactive decay) model combined with two afterglow components. The right-hand panel shows the same data with the supernova component (dashed lines) and the afterglow components (solid lines), highlighting how the thermal component begins to dominate after $\sim 20$ days.}
    \label{fig:opticalfit}
\end{figure*}

In Figure~\ref{fig:aft_model}, left panel, we show model light curves that are consistent with the behavior seen within the data.
At $<6$~days (Epoch~I), we describe the afterglow with a top-hat jet viewed on-axis with an opening angle of $\theta_j \simeq 15\fdg0$, an isotropic equivalent kinetic energy $6.3\times 10^{51}$~erg, an initial Lorentz factor $\Gamma_0 = 100$, and 
$p = 1.8$ (the microphysical parameters were fixed).\footnote{The fraction of energy given to electrons $\varepsilon_e = 0.1$, the fraction of energy given to magnetic fields $\varepsilon_B = 0.01$, the participation fraction of electrons in the synchrotron process $\Xi_N = 0.1$, and the ambient density is assumed to be uniform with a number density $n = 10$~cm$^{-3}$ throughout. We use \texttt{Redback} afterglow models \citep{Sarin2024}, \texttt{tophat\_redback} and \texttt{twocomponent\_redback} (with a hollow core) for the afterglow lightcurves \citep[see][for details]{Lamb2017, Lamb2018b}.}

To describe the light curve at $> 8$~days (Epoch II), we require two components: $i)$ a late epoch afterglow re-energized copy of the Epoch~I model (i.e., afterglow transitions) via some energy injection process, with an isotropic kinetic energy $E_{\rm k,iso} = 2\times 10^{52}$~erg, and $p = 2.0$, all else as previous (this component is shown as a narrow solid line in Fig.~\ref{fig:aft_model}, left panel); and
$ii)$ an additional component is required to describe the radio and late X-ray data.
To model this, we invoke a sheath, or two-component structure that encases the jet core (this component is shown as a dashed line in Fig.~\ref{fig:aft_model}, left panel).
This sheath extends from the edge of the jet core, at $\theta_j = 15\fdg0$, to the outer sheath at $\theta_s = 20\fdg0$.
The isotropic equivalent kinetic energy within the sheath is $E_{\rm k,iso} = 5\times 10^{52}$~erg, the Lorentz factor of the sheath material is $\Gamma_{0,s} = 3.5$, and the participation fraction is $\Xi_{N,s} = 0.2$.
The total light curve, the sum of the early or late core, respectively, and the sheath, is shown as a thick solid line.

The model light curve fails to reproduce the late X-ray emission at Epoch~III.
Additionally, the complexity of the radio variability is not exactly reproduced; however, some of this variability may be due to scintillation, which is not modeled.
The late optical and infrared data during Epoch~III appear in excess of the afterglow model -- this is expected as the optical and NIR are dominated by an unmodeled thermal component at this time.

\subsection{Epoch III: Supernova modeling}

Drawing motivation from the discussion above (see \S\ref{sec:pheno}), we now consider the origin of the thermal component that begins to dominate the optical emission during Epoch~III. To simplify our modeling, we fix the non-thermal emission components as described above and explore the optical light curve post 15~days as a combination of a two non-thermal components (dominating the emission during Epochs~I and II) and one thermal component\footnote{We use {\sc Redback}~\citep{Sarin2024} implementations of all models and fit for the thermal component using the {\sc pymultinest}~\citep{Feroz2009, Buchner2016} sampler via {\sc Bilby}~\citep{Ashton2019}.}. We first consider TDE-like emission, although we note that no TDE has ever been associated with the highly relativistic outflows \citep[i.e., $\Gamma\approx 100$; e.g., AT2022cmc and Sw~J1644+57 are modeled to have had $\Gamma\approx 10$;][]{Andreoni2022,Bloom2011}, which we require to explain the rest of the light curve, as discussed above. This includes both jetted TDE systems where the jet was launched promptly and systems with delayed relativistic outflows relative to optical \citep{Cendes2024}.
We consider two models for optical emission from a TDE: $i)$ the optical luminosity is a function of the fall-back rate onto the BH \citep{Guillochon2013, Mockler2019}; and $ii)$ the optical emission is generated via cooling emission of a hot pressure-supported envelope of the disrupted material \citep{Metzger2022, Sarin2024_cooling}. 
In general, similar to the models shown in Fig.~\ref{fig:IMBH_TDE}, we find that TDE models (in the relevant BH mass range) do not sufficiently add to the afterglow light curve to explain the thermal component. Moreover, tuning parameters such as the efficiency of the BH to resolve this lead to thermal light curves that last too long, inconsistent with the late-time behavior seen in EP~241021a.

Next, we consider a supernova origin for the thermal component, which, as we discussed, appears to be the likely origin of the thermal component during Epoch~III, consistent with the time scale, luminosity, and colors expected for a Type Ic-BL SN. We attempt four different fits to the data (with the exact setup as above). First, we consider emission from radioactive decay of ${}^{56}$Ni following a one-zone model \citep{Arnett1980, Nadyozhin1994}. Our fits to the overall optical light curve from 15 days is shown in Fig.~\ref{fig:opticalfit} (left panel), which shows great agreement with the data, and how the SN component (right panel, dashed lines) begins to dominate over the declining afterglow component (right panel, solid lines) at $t \gtrsim 18$ days, consistent with our analysis from the SED. However, while this model broadly works at describing the data, the estimated parameters are in tension with expectations for other Type Ic-BL SNe. In particular, we recover an estimated ejecta mass of $4.7^{+2.6}_{-1.7}M_{\odot}$ with a nickel fraction of $0.5^{+0.3}_{-0.2}$ and extremely high velocities of $\sim 150,000$~km~s$^{-1}$, which are all inconsistent with constraints from other Type Ic \citep[e.g.,][]{Srinivasaragavan2024}.

This discrepancy between our inferred parameters and those of other SNe could be due to several reasons, such as our analysis fixing the components that describe the complex nature of the afterglow. We also perform fits with arbitrarily scaled and shifted SEDs of SN~1998bw, which fail to adequately describe the full light curve. This could raise doubts regarding a supernova origin for the thermal component. However, we believe that the high energetics and fast duration of the thermal component in EP~241021a compared to other Type Ic-BL SNe (see Fig.~\ref{fig:dura_peak}) suggest another energy source is at play. Such an additional component could also be responsible for the re-brightening in X-rays and optical. To further investigate this scenario, we perform additional fits with models invoking additional energy in a SN from fallback accretion onto a stellar-mass BH~\citep{Sarin2024}, a millisecond magnetar~\citep{Sarin2022, Omand2024}, and circumstellar-medium (CSM)-interaction \citep{Chatzopoulos2012}. We find that all three models provide good fits to the data (with a preference for the magnetar model based on the Bayes factor). Furthermore, the additional energy input from either source also alleviates the concern about our ejecta parameters, for example, across all three models we infer ${}^{56}\rm{Ni} \approx 0.15$, consistent with expectations and prior constraints from Type Ic SNe \citep{Srinivasaragavan2024}. Ultimately, which of these interpretations is correct and precise estimates of parameters relies on much more detailed modeling (including the afterglow component in the fit). However, we also stress that the lack of detailed optical data will likely prevent any strong conclusions, as these models are intrinsically uncertain and degenerate. The broad agreement of the data during Epoch~III with a combination of two afterglows, SN component plus energy injection, suggests the progenitor of EP~241021a is similar to a few other EP FXTs \citep[e.g.,][]{van_Dalen2024,Rastinejad2025,Eyles_Ferris2025, Srinivasaragavan2025}, and these events are mostly related to the collapsar phenomena. The exceptional nature of the likely SN in EP~241021a points to an additional source of energy besides radioactive decay from ${}^{56}\rm{Ni}$. 

\citet{Busmann2025} also modeled their light curve data for EP~241021a with radioactive decay models and included an additional energy input from CSM interaction and magnetar models. Only the former two models are consistent with the implemented physics, as the models we use here, but our inferred parameters are considerably different. For example, they infer an ejecta mass of $9.9 M_{\odot}$ with $\sim 50\%$ ${}^{56}\rm{Ni}$ for the radioactive-decay only model, entirely inconsistent with any Type Ic supernova~\citep{Srinivasaragavan2024}. Similarly, for the CSM-interaction with radioactive decay model, they infer a CSM-mass of $\approx 95\pm 60 M_{\odot}$, implying an extremely massive progenitor. These constraints and discrepancies are most likely driven by the assumptions of the non-thermal afterglow component. We note that our magnetar model is significantly different from the model used in their work, including, for example, losses due to work done by the magnetar wind \citep[see][for details]{Sarin2022, Omand2024} that is ignored in their implementation, so our results are not comparable.

\section{Conclusions}\label{sec:conclusion}

We have presented a detailed multi-wavelength study of the fast X-ray transient EP~241021a, discovered on 2024 October 21 by the EP-WXT instrument aboard the Einstein Probe. With a redshift of $z = 0.7485$ and a peak X-ray luminosity of $\sim2\times10^{48}$erg~s$^{-1}$, EP~241021a exhibited a rich and complex evolution across the X-ray, optical, NIR, and radio bands over a period of more than $\sim$100 days.

The optical light curve is characterized by a distinct three-phase morphology (see Fig.~\ref{fig:lc_fit}), captured through a smooth triple power-law fit: (I) an initial decay with power-law index $\alpha \approx -0.7$ lasting until day 5.6; (II) a sharp re-brightening peaking at $\sim7.7$ days with $\alpha \approx +3.6$, followed by a steeper decay than in the previous phase with a slope of $\alpha \approx -1.3$; and (III) a mild re-brightening around day $\sim$19, followed by a slower decline. This light curve evolution is not typical of standard GRB afterglows and suggests the presence of multiple physical components contributing to the observed emission.

During the first $\sim$20 days, the optical and X-ray SEDs are both well described by a single non-thermal power law (see Fig.~\ref{fig:SED}), with spectral indices consistent with optically thin synchrotron emission ($\beta_{\rm op} \approx -1.1$ and $\beta_{\rm ox} \approx -1.0$), indicating that the optical and X-ray emission likely originate from the same spectral component. However, beyond day $\sim$20, the optical SED becomes significantly steeper and inconsistent with the X-ray data, reaching a spectral slope of $\beta_{\rm op} \approx -3.1$ (see Fig.~\ref{fig:SED}). Concurrently, the optical source rebrightened, reaching an absolute magnitude of $M_r \approx -20.1$~mag (see Fig.~\ref{fig:lc_comparison}).

Spectroscopically, the transient displayed a relatively stable optical continuum with a spectral slope of $\beta_{\rm spec} \approx -1.2$ until day $\sim$19 (see Fig.~\ref{fig:spec_evol}), confirming the photometric evidence of a non-thermal emission component. However, subsequent spectroscopic epochs reveal a different behavior than in earlier phases, especially our MUSE spectrum shows an even steeper spectral index of $\beta_{\rm spec} \approx -2.73$. This spectral shape is interpreted as due to the emergence of a new emission component. However, no broad absorption lines characteristic of type Ic-BL supernovae were detected.

The inclusion of contemporaneous radio, NIR, optical, and X-ray data from follow-up campaigns \citep[including those presented before in the literature;][]{Gianfagna2025,Yadav2025,Busmann2025,Shu2025} enables broadband SED fitting (see Fig.~\ref{fig:SED_radio_xray}). 
Both from analyzing our optical data as well as data in the literature, we conclude that a non-thermal origin is required for the early-time emission. However, beyond day $\sim$20, an additional blackbody component is required to explain the optical SED, alongside the non-thermal contribution. The onset of the re-brightening after day $\gtrsim$20, in combination with the steepening of the optical spectral slope and the evolving SED, is well explained by the emergence of a thermal component. Indeed, the host galaxy does not contribute significantly to the photometry of the transient until after day $\approx70$.
This is interpreted as the onset of a supernova, supported by its position in the peak absolute magnitude versus duration parameter space (see Fig.~\ref{fig:dura_peak}), its color evolution (see Fig.~\ref{fig:colors}), and the subtracted-afterglow MUSE spectrum at late times (see Fig.~\ref{fig:type_Ic}). The presence of this thermal component and its consistency with a type Ic-BL supernova signature suggest that the progenitor of this X-ray transient detected by EP might be a collapsar. Our inability to detect typical type Ic BL SN  features might be caused by the presence of an additional spectral component reducing the equivalent width of the lines, in combination with the relatively high redshift of the event.

We modeled the complex light curve of EP~241021a as the superposition of some distinct afterglow components, each representing a separate emission region or energy release episode. The first component, responsible for the early-time decay ($\lesssim$6 days), corresponds to the standard forward-shock afterglow produced by the interaction of a relativistic outflow with the ISM. This model assumes a top-hat jet viewed on-axis, with an opening angle of $\theta_j \approx 15^\circ$, an isotropic-equivalent kinetic energy of $6.3\times10^{51}$~erg, an initial Lorentz factor $\Gamma_0 = 100$, and a power-law index of $p = 1.8$ for the shock-accelerated electrons. This component follows a smooth decline with temporal and spectral indices consistent with synchrotron emission in the slow-cooling regime.
To reproduce the light curve beyond $\sim$8 days, two additional components are required. The second component shares characteristics with the first, but incorporates an energy injection episode that increases the kinetic energy by a factor of $\approx$2, while maintaining similar microphysical parameters. A third component, needed to account for the radio and late-time X-ray data, involves a sheath or two-component jet structure surrounding the jet core. In addition, to explain the late-time optical behavior, a supernova component must be included, consistent with the spectro-photometric evidence. This multi-component framework reinforces the interpretation of EP~241021a as a hybrid transient, exhibiting both relativistic jet dynamics and late-time energy sources such as a supernova and potential prolonged central engine activity.

EP~241021a exemplifies the diverse and evolving population of extragalactic fast X-ray transients now being uncovered by the Einstein Probe satellite. Its rich multi-phase light curve, delayed spectral transition, and ambiguous classification challenge existing models and highlight the importance of early, coordinated, and long-term follow-up across the electromagnetic spectrum. 

\section*{Acknowledgements}

In memory of Miguel Angel V\'asquez Lasso, lovely grandfather.
J.Q.V., P.G.J., J.N.D.D., J.S.S., M.E.R., and A.P.C.H. are supported by the European Union (ERC, Starstruck, 101095973, PI Jonker). Views and opinions expressed are, however, those of the author(s) only and do not necessarily reflect those of the European Union or the European Research Council Executive Agency. Neither the European Union nor the granting authority can be held responsible for them. 
J.Q.V. additionally acknowledges support by the IAU-Gruber
foundation.

D.M.S. and M.A.P.T. acknowledge support by the Spanish Ministry of Science via the Plan de Generación de conocimiento PID2020-120323GB-I00. DMS also acknowledges support via a Ramon y Cajal Fellowship RYC2023-044941. 
N.S acknowledges support from the Knut and Alice Wallenberg Foundation through the ``Gravity Meets Light" project and the research environment grant ``Gravitational Radiation and Electromagnetic Astrophysical Transients'' (GREAT) funded by the Swedish Research Council (VR) under Dnr 2016-06012.
GPL is supported by a Royal Society Dorothy Hodgkin Fellowship (grant Nos. DHF-R1-221175 and DHF-ERE-221005).
P.T.O.B acknowledges support from UKRI grant ST/W000857/1. F.E.B. acknowledges support from ANID-Chile BASAL CATA FB210003, FONDECYT Regular 1241005, and Millennium Science Initiative, AIM23-0001. 
The work of D. S. was carried out at the Jet Propulsion Laboratory, California Institute of Technology, under a contract with the National Aeronautics and Space Administration (80NM0018D0004).

Based on observations obtained at the international Gemini Observatory (Program IDs GN-2024B-Q-107, GS-2024B-Q-105, GS-2025A-Q-107), a program of NOIRLab, which is managed by the Association of Universities for Research in Astronomy (AURA) under a cooperative agreement with the National Science Foundation on behalf of the Gemini Observatory partnership: the National Science Foundation (United States), National Research Council (Canada), Agencia Nacional de Investigaci\'{o}n y Desarrollo (Chile), Ministerio de Ciencia, Tecnolog\'{i}a e Innovaci\'{o}n (Argentina), Minist\'{e}rio da Ci\^{e}ncia, Tecnologia, Inova\c{c}\~{o}es e Comunica\c{c}\~{o}es (Brazil), and Korea Astronomy and Space Science Institute (Republic of Korea). Data was processed using the Gemini DRAGONS (Data Reduction for Astronomy from Gemini Observatory North and South) package.

Data for this paper has in part been obtained under the International Time Programme of the CCI (International Scientific Committee of the Observatorios de Canarias
of the IAC) with the NOT and GTC operated on the island of La Palma by the Roque de los Muchachos.

The Legacy Surveys consist of three individual and complementary projects: the Dark Energy Camera Legacy Survey (DECaLS; Proposal ID \#2014B-0404; PIs: David Schlegel and Arjun Dey), the Beijing-Arizona Sky Survey (BASS; NOAO Prop. ID \#2015A-0801; PIs: Zhou Xu and Xiaohui Fan), and the Mayall $z$-band Legacy Survey (MzLS; Prop. ID \#2016A-0453; PI: Arjun Dey). DECaLS, BASS and MzLS together include data obtained, respectively, at the Blanco telescope, Cerro Tololo Inter-American Observatory, NSF’s NOIRLab; the Bok telescope, Steward Observatory, University of Arizona; and the Mayall telescope, Kitt Peak National Observatory, NOIRLab. Pipeline processing and analyses of the data were supported by NOIRLab and the Lawrence Berkeley National Laboratory (LBNL). The Legacy Surveys project is honored to be permitted to conduct astronomical research on Iolkam Du\'ag (Kitt Peak), a mountain with particular significance to the Tohono O\'odham Nation.\\

\section*{Data Availability}

The data will be available and made public once the paper is published.



\bibliographystyle{mnras}
\bibliography{Quirola-Vasquez} 




\appendix

\section{Comparison with other transients}

\begin{figure}
    \centering
    \includegraphics[scale=0.7]{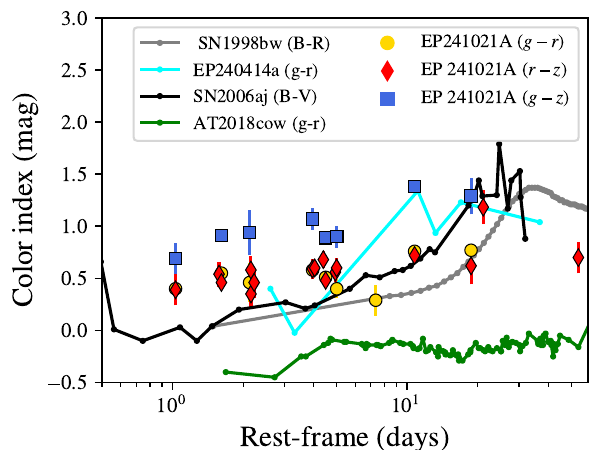}
    \vspace{-0.2cm}
    \caption{Color evolution of EP~241021a, compared to that of the transients AT~2018cow, SN~1998bw, SN~2006aj, and EP~240414a \citep{Soderberg2006,Prentice2018,Clocchiatti2011,van_Dalen2024}. The blue, red, and orange markers depict the respective $g-z$, $r-z$, and $g-r$ colors of EP~241021a.}
    \label{fig:colors}
\end{figure}

\begin{figure}
    \centering
    \includegraphics[scale=0.85]{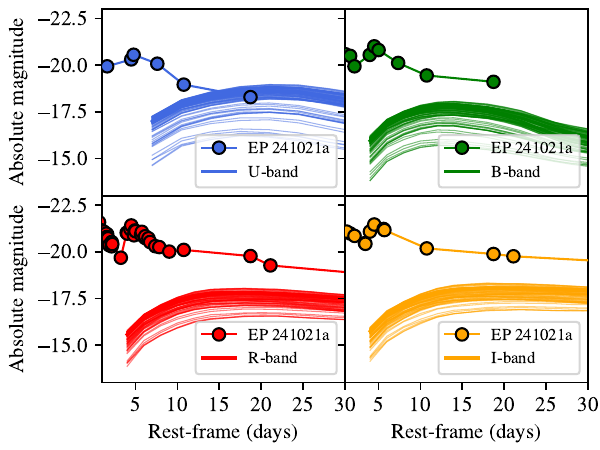}
    \vspace{-0.2cm}
    \caption{Optical light curves of EP~241021a compared to optical thermonuclear transients from hydrodynamic simulation (considering a width range of orientations) of tidal disruption events of white dwarfs from \citep{MacLeod2016}. The optical light curves remain brighter than the simulations during the whole emission.}
    \label{fig:IMBH_TDE}
\end{figure}

As we mentioned above, the comparison with well-known transients is essential to discover the progenitor of EP~241021a.
Figure~\ref{fig:colors} depicts the color evolution of EP~241021a, and a comparison with transients such as SN~1998bw \citep{Clocchiatti2011}, AT2018cow \citep{Prentice2018}, SN~2006aj \citep{Soderberg2006}, and EP~240414a \citep{van_Dalen2024}.
Meanwhile, Fig.~\ref{fig:IMBH_TDE} shows the optical multi-band light curves of EP~241021a and the hydrodynamical simulation developed by \citet{MacLeod2016} for the tidal disruption of WD by IMBH.

\newpage

\begin{figure}
    \centering
    \includegraphics[scale=0.5]{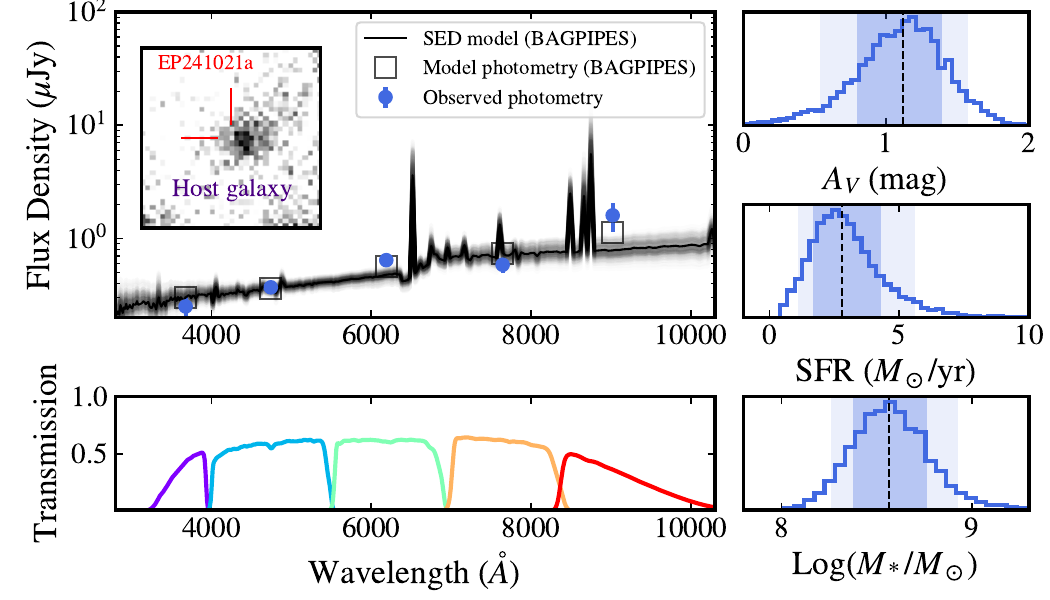}
    \vspace{-0.5cm}
    \caption{{\it Main panel:} Posterior SED models for the host galaxy (observed by HiPERCAM using $ugriz$ bands) of EP~241021a obtained from \texttt{Bagpipes}. {\it Bottom panel:} The relative transmission functions of the different filters are used in the fitting process. {Right panels:} From top to bottom: the posterior distribution of the dust attenuation, the star-formation rate (SFR), and the stellar mass. The blue hue gradient depicts the 68\% and 90\% confidence levels, respectively. The inset figure in the top main panel shows the image in $r$-band taken by HiPERCAM $\sim304$~days after the trigger of the host galaxy of the transient. The red lines mark the position of EP~241021a.}
    \label{fig:host_sed}
\end{figure}

\begin{table*}
    \centering
    \scalebox{0.9}{
    \begin{tabular}{lllccccc}
    \hline\hline
    Telescope & Instrument & Date (UT) & Days since trigger & Exposure time & Filter & AB magnitude & Ref. \\ 
    (1) & (2) & (3) & (4) & (5) & (6) & (7) & (8) \\ \hline
NOT & ALFOSC & 2024-10-22 23:41:02 & 1.773   & 5$\times$200 & $z$ & 21.61$\pm$0.11 & This work \\
NOT & ALFOSC & 2024-10-23 03:01:58 & 1.91252 & 5$\times$300 & $r$ & 21.95$\pm$0.06 & This work \\
NOT & ALFOSC & 2024-10-23 23:19:27 & 2.75801 & 5$\times$360 & $r$ & 21.99$\pm$0.05 & This work \\
NOT & ALFOSC & 2024-10-24 00:48:23 & 2.81976 & 5$\times$300 & $z$ & 21.55$\pm$0.11 & This work \\
GTC & HiPERCAM & 2024-10-24 00:43:21 & 2.8232 & 40$\times$30 & $u$ & 22.85$\pm$0.04 & This work \\
GTC & HiPERCAM & 2024-10-24 00:43:21 & 2.8232 & 40$\times$30 & $g$ & 22.29$\pm$0.02 & This work \\
GTC & HiPERCAM & 2024-10-24 00:43:21 & 2.8232 & 40$\times$30 & $r$ & 21.84$\pm$0.02 & This work \\
GTC & HiPERCAM & 2024-10-24 00:43:21 & 2.8232 & 40$\times$30 & $i$ & 21.80$\pm$0.02 & This work \\
GTC & HiPERCAM & 2024-10-24 00:43:21 & 2.8232 & 40$\times$30 & $z$ & 21.48$\pm$0.03 & This work \\
\emph{Swift} & UVOT & 2024-10-24 20:53:07 & 3.7236 & 2545.8 & $V$ & $>$20.69 & This work \\
NOT & ALFOSC & 2024-10-24 23:41:21 & 3.77322 & 5$\times$300 & $r$ & 22.26$\pm$0.05 & This work \\
NOT & ALFOSC & 2024-10-25 00:06:08 & 3.79042 & 5$\times$300 & $z$ & 22.01$\pm$0.12 & This work \\
GS & GMOS-acq & 2024-10-25 2:32:22 & 3.89231 & 1$\times$60 & $r$ & 22.37$\pm$0.08 & This work \\
Keck & MOSFIRE & 2024-10-25 9:57:55 & 4.201377 & 5$\times$60 & $J$ & 21.86$\pm$0.05 & This work \\
Keck & MOSFIRE & 2024-10-25 10:06:50 & 4.207569 & 6$\times$30 & $H$ & 21.76$\pm$0.12 & This work \\
Keck & MOSFIRE & 2024-10-25 10:19:38 & 4.216458 & 9$\times$60 & $K_s$ & 21.45$\pm$0.06 & This work \\
\emph{Swift} & UVOT & 2024-10-25 15:30:30 & 4.47832 & 1347.82 & $UVM2$ & $>$22.49 & This work \\
SOAR & Goodman & 2024-10-27 03:08:35 & 5.917132 & 15$\times$100 & $i$ & 22.32$\pm$0.11 & This work \\
GS & GMOS & 2024-10-28 02:39:48 & 6.89714 & 4$\times$180 & $r$ & 21.76$\pm$0.01 & This work \\
GS & GMOS & 2024-10-28 02:52:48 & 6.91 & 4$\times$180 & $g$ & 22.24$\pm$0.02 & This work \\
GS & GMOS-acq & 2024-10-28 3:26:12 & 6.929699 & 1$\times$60 & $z$ & 21.27$\pm$0.10 & This work \\
GS & GMOS-acq & 2024-10-28 3:51:46 & 6.94745 & 1$\times$60 & $i$ & 21.7$\pm$0.05 & This work \\
GTC & HiPERCAM & 2024-10-29 01:31:00.9 & 7.8563 & 40$\times$30 & $u$ & 22.48$\pm$0.02 & This work \\
GTC & HiPERCAM & 2024-10-29 01:31:00.9 & 7.8563 & 40$\times$30 & $g$ & 21.78$\pm$0.01 & This work \\
GTC & HiPERCAM & 2024-10-29 01:31:00.9 & 7.8563 & 40$\times$30 & $r$ & 21.37$\pm$0.01 & This work \\
GTC & HiPERCAM & 2024-10-29 01:31:00.9 & 7.8563 & 40$\times$30 & $i$ & 21.32$\pm$0.01 & This work \\
GTC & HiPERCAM & 2024-10-29 01:31:00.9 & 7.8563 & 40$\times$30 & $z$ & 20.99$\pm$0.02 & This work \\
\emph{Swift} & UVOT & 2024-10-29 09:32:49 & 8.3458 & 4289.06 & $U$ & 22.23$\pm$0.24 & This work \\
NOT & ALFOSC & 2024-10-29 23:03:22 & 8.74684 & 3$\times$300 & $r$ & 21.68$\pm$0.06 & This work \\
NOT & ALFOSC & 2024-10-29 23:19:58 & 8.75836 & 3$\times$300 & $g$ & 21.98$\pm$0.07 & This work \\
NOT & ALFOSC & 2024-10-29 23:42:02 & 8.77369 & 5$\times$300 & $z$ & 21.18$\pm$0.08 & This work \\
GTC  & OSIRIS-acq & 2024-10-31 00:36:59 & 9.81185 & 1$\times$150 & $i$ & 21.56$\pm$0.05 & This work \\
SOAR & Goodman & 2024-10-31 02:38:58 & 9.89648 & 20$\times$60 & $r$ & 21.87$\pm$0.08 & This work \\
SOAR & Goodman & 2024-10-31 04:30:17 & 9.9739 & 20$\times$60 & $i$ & 21.62$\pm$0.09 & This work \\
GS & GMOS-acq & 2024-10-31 6:16:25 & 10.04859 & 1$\times$180 & $r$ & 21.71$\pm$0.02 & This work \\
NOT & ALFOSC & 2024-10-31 23:18:48 & 10.75756 & 4$\times$300 & $r$ & 22.04$\pm$0.05 & This work \\
NOT & ALFOSC & 2024-11-1 23:55:37  & 11.78311 & 4$\times$300 & $r$ & 22.26$\pm$0.26 & This work \\
SOAR & Goodman & 2024-11-3 01:18:53 & 12.8409 & 25$\times$60 & $g$ & 22.67$\pm$0.11 & This work \\
SOAR & Goodman & 2024-11-3 01:56:03 & 12.86676 & 25$\times$60 & $r$ & 22.48$\pm$0.10 & This work \\
\emph{Swift} & UVOT & 2024-11-03 02:25:38 & 13.2881 & 3059.02 & $U$ & 22.71$\pm$0.43 & This work \\
\emph{Swift} & UVOT & 2024-11-05 20:19:57 & 15.6522 & 985.45 & $U$ & $>$21.87 & This work \\
NOT & ALFOSC & 2024-11-5 23:57:39  & 15.78453 & 4$\times$300 & $r$ & 22.77$\pm$0.06 & This work \\
\emph{Swift} & UVOT & 2024-11-07 02:11:57 & 17.1315 & 3107.47 & $U$ & $>$22.52 & This work \\
GTC & HiPERCAM & 2024-11-09 00:13:48 & 18.8027 & 40$\times$30 & $u$ & 23.83$\pm$0.09 & This work \\
GTC & HiPERCAM & 2024-11-09 00:13:48 & 18.8027 & 40$\times$30 & $g$ & 23.34$\pm$0.03 & This work \\
GTC & HiPERCAM & 2024-11-09 00:13:48 & 18.8027 & 40$\times$30 & $r$ & 22.68$\pm$0.04 & This work \\
GTC & HiPERCAM & 2024-11-09 00:13:48 & 18.8027 & 40$\times$30 & $i$ & 22.60$\pm$0.06 & This work \\
GTC & HiPERCAM & 2024-11-09 00:13:48 & 18.8027 & 40$\times$30 & $z$ & 22.06$\pm$0.05 & This work \\
GTC & EMIR & 2024-11-10 23:35:08 & 20.7688 & 260$\times$5 & $H$ & 22.39$\pm$0.09 & This work \\
\emph{Swift} & UVOT & 2024-11-15 19:08:48 & 25.8468 & 1676.24 & $U$ & $>$22.22 & This work \\
SOAR & Goodman & 2024-11-18 03:12:09 & 27.92751 & 15$\times$80 & $z$ & 22.48$\pm$0.32 & This work \\
SOAR & Goodman & 2024-11-18 04:38:14 & 27.98729 & 15$\times$80 & $r$ & $>$22.65 & This work \\
GTC & HiPERCAM & 2024-11-22 22:37:41 & 32.7359 & 40$\times$30 & $u$ & 24.50$\pm$0.12 & This work \\
GTC & HiPERCAM & 2024-11-22 22:37:41 & 32.7359 & 40$\times$30 & $g$ & 23.68$\pm$0.04 & This work \\
GTC & HiPERCAM & 2024-11-22 22:37:41 & 32.7359 & 40$\times$30 & $r$ & 23.01$\pm$0.05 & This work \\
GTC & HiPERCAM & 2024-11-22 22:37:41 & 32.7359 & 40$\times$30 & $i$ & 22.90$\pm$0.07 & This work \\
GTC & HiPERCAM & 2024-11-22 22:37:41 & 32.7359 & 40$\times$30 & $z$ & 22.49$\pm$0.17 & This work \\
VLT & MUSE & 2024-11-27 01:52:05 & 36.8956 & 3$\times$700 & $r$ & 23.51$\pm$0.11 & This work \\
VLT & MUSE & 2024-11-27 01:52:05 & 36.8956 & 3$\times$700 & $i$ & 23.02$\pm$0.11 & This work \\
VLT & MUSE & 2024-11-27 01:52:05 & 36.8956 & 3$\times$700 & $z$ & 22.43$\pm$0.12 & This work \\
GS & GMOS & 2024-12-29 01:23:20 & 68.86419 & 10$\times$180 & $r$ & 24.24$\pm$0.05 & This work \\
GS & GMOS & 2025-01-22 01:35:50	& 92.8616 & 12$\times$80 & $z$ & 23.94$\pm$0.22 & This work \\
GS & GMOS & 2025-01-23 00:57:01 & 93.83952 & 12$\times$180 & $r$ & 24.52$\pm$0.06 & This work \\
GTC & HiPERCAM & 2025-02-01 20:53:20.68 & 103.663 & 40$\times$30 & $g$ & $>$24.69 & This work \\
GTC & HiPERCAM & 2025-02-01 20:53:20.68 & 103.663 & 40$\times$30 & $r$ & $>$24.15 & This work \\
GTC & HiPERCAM & 2025-02-01 20:53:20.68 & 103.663 & 40$\times$30 & $i$ & 23.95$\pm$0.39 & This work \\
GTC & HiPERCAM & 2025-02-01 20:53:20.68 & 103.663 & 40$\times$30 & $z$ & $>$24.18 & This work \\
\hline
    \end{tabular}
    }
    \caption{Photometry obtained with various ground-based telescopes for this work. UL in the AB magnitude column stands for the 3$\sigma$ upper limit. \emph{Columns 1 and 2:} Telescope and instrument per observation, respectively. \emph{Column 3 and 4:} start date of the observation and days after the X-ray trigger, respectively. \emph{Columns 5 and 6:} exposure time and filter per observation, respectively. \emph{Column 7:} magnitude and uncertainty in AB system. The photometry was not corrected for Galactic extinction, nor for the host contribution.}
    \label{tab:photometry}
\end{table*}

\begin{table*}
    \centering
    \scalebox{0.85}{
    \begin{tabular}{lllccccc}
    \hline\hline
    Telescope & Instrument & Date (UT) & Days since trigger & Exposure time & Filter & AB magnitude & Ref. \\ 
    (1) & (2) & (3) & (4) & (5) & (6) & (7) & (8) \\ \hline
    KAIT & -- & 24-10-22 08:24:00	& 1.14 & $R$ & 70$\times$60 & 21.2$\pm$0.2 & \cite{Zheng2024GCNa}	\\
    DFOT & -- & 2024-10-22 16:48:07 & 1.49 & $R$ & 12$\times$300 & 21.62$\pm$0.1 & \cite{Ror2024GCN} \\
    Liverpool Telescope	 & -- &	24-10-23 00:00:00 & 1.8	& $g$ &	-- & 22.2$\pm$0.1 & \cite{Li2024GCNa} \\	
    Liverpool Telescope	& -- & 24-10-23 00:00:00 & 1.8	& $r$ & -- & 21.9$\pm$0.1 & \cite{Li2024GCNa} \\	
    1m Las Cumbres observatory & -- & 24-10-23 07:12:00 & 2.1 & $i$ & -- & 21.7$\pm$0.1	& \cite{Li2024GCNb} \\	
    1m Las Cumbres observatory & -- & 24-10-23 07:12:00 & 2.1	& $r$ &	-- & 21.9$\pm$0.1 &	\cite{Li2024GCNb} \\	
    KAIT & -- & 24-10-23 07:55:12	& 2.12 & $R$ & 70$\times$60 & 21.7$\pm$0.3 & \cite{Zheng2024GCNa}	\\
    2.16m Xinglong & -- & 2024-10-23 15:57:29 & 2.42 & $V$ & 3$\times$600 &	22.5$\pm$0.5 & \cite{Jin2024GCNa} \\
    1-m SAO RAS	& CCD-photometer & 2024-10-23 19:27:12 & 2.61 & $R$ & 9$\times$300 & 21.81$\pm$0.06 & \cite{Moskvitin2024GCN} \\ 
    VLT & FORS2-acq & 24-10-24 05:29:45 & 2.998 & $r$ & -- & 22.06$\pm$0.05 & \cite{Pugliese2024GCN} \\
    Liverpool Telescope & -- & 2024-10-24 23:07:43 & 3.25 & $r$ & 6$\times$150 & 22.43$\pm$0.14 & \cite{Bochenek2024GCN} \\
    Liverpool Telescope & -- & 2024-10-24 22:28:08.2 & 3.72 & $g$ & 3$\times$300 & 22.85$\pm$0.18 & \cite{Kumar2024GCN} \\
    Liverpool Telescope & -- & 2024-10-24 22:44:16.8 & 3.73 & $i$ & 3$\times$300 & 21.93$\pm$0.15 & \cite{Kumar2024GCN} \\
    Fraunhofer Telescope & 3KK & 2024-10-24 02:12:17 & 3.77 & $r$ & 40$\times$180 & 22.49$\pm$0.05 & \cite{Busmann2024GCN} \\
    1-m SAO RAS	& CCD-photometer & 2024-10-26 20:39:47 & 5.65 & $R$ & 12$\times$300 & 23.10$\pm$0.18 & \cite{Moskvitin2024GCNb} \\
    SOAR & Goodman & 2024-10-27 04:56:18 & 6.004 & $i$ & 4$\times$300 & 22.36$\pm$0.07 & \cite{Freeburn2024GCN} \\
    Fraunhofer Telescope & 3KK & 2024-10-28 21:08:46 & 7.67 & $J$ & 5092 & $20.79\pm0.10$ & \cite{Busmann2025} \\
    1-m SAO RAS	& CCD-photometer & 2024-10-28 19:41:42 & 7.68 & $R$ & 15$\times$300 & 21.57$\pm$0.08 & \cite{Moskvitin2024GCNc} \\
    Mephisto & -- & 2024-10-29 15:15:40 & 8.42 & $r$ & 15$\times$300 & 21.9$\pm$0.2 & \cite{Pan2024GCN} \\
    1-m SAO RAS	& CCD-photometer & 2024-10-29 19:55:32 & 8.64 & $R$ & 12$\times$300 & 21.64$\pm$0.05 & \cite{Moskvitin2024GCNc} \\
    T193cm telescope & MISTRAL & 2024-10-31 21:11:05 & 10.7 & $r$ & 1$\times$300s + 5$\times$600 & 22.01$\pm$0.05 & \cite{Schneider2024GCN} \\
    Liverpool Telescope & -- & 2024-10-31 23:32:50 & 10.77 & $r$ & 6$\times$200 & 21.95$\pm$0.08 & \cite{Bochenek2024GCNb} \\
    1m LOT & -- & 2024-11-01 15:44:21 & 11.44 & $r$ & 6$\times$300 & 22.06$\pm$0.33 & GCN 38042 \\
    T193cm telescope & MISTRAL & 2024-11-03 20:29:00 & 13.7 & $r$ & 12$\times$600s & 22.53$\pm$0.12 & GCN 38071 \\
    Fraunhofer Telescope & 3KK & 2024-11-07 21:53:27 & 17.70 & $J$ & 6790 & 21.93$\pm$0.16 & \cite{Busmann2025} \\
    GTC & EMIR & -- & 23.3 & $J$ & 350 & 21.58$\pm$0.37 & \cite{Gianfagna2025} \\
    GTC & EMIR & -- & 23.3 & $H$ & 462 & 21.98$\pm$0.39 & \cite{Gianfagna2025} \\
    GTC & EMIR & -- & 23.3 & $K_s$ & 588 & 21.73$\pm$0.36 & \cite{Gianfagna2025} \\
    Fraunhofer Telescope & 3KK & 2024-11-16 18:12:12 & 26.54 & $r$ & 4140 & 23.3$\pm$0.4 & \cite{Busmann2025} \\
    Fraunhofer Telescope & 3KK & 2024-11-16 18:12:12 & 26.54 & $z$ & 3960 & 22.7$\pm$0.4 & \cite{Busmann2025} \\
    Fraunhofer Telescope & 3KK & 2024-11-24 20:59:37 & 34.66 & $J$ & 8703 & $>$21.7 & \cite{Busmann2025} \\
    Fraunhofer Telescope & 3KK & 2024-11-27 19:32:13 & 37.60 & $J$ & 3565 & $>$21.9 & \cite{Busmann2025} \\    
    \hline
    \end{tabular}
    }
    \caption{Similar to Table~\ref{tab:photometry}, but considering only the photometry that is publicly available for EP~241021a obtained with various ground-based telescopes.}
    \label{tab:photometry_gcn}
\end{table*}

\begin{table*}
    \centering
    \scalebox{0.9}{
    \begin{tabular}{lllcccccc}
    \hline\hline
        Telescope & Instrument & Date (UT) & Days since trigger & Exp. time (ks) & Band (keV) & $\Gamma$ & Flux (erg~s$^{-1}$~cm$^{-2}$) & Reference \\
        (1) & (2) & (3) & (4) & (5) & (6) & (7) & (8) & (9) \\
        \hline
         Einstein Probe & WXT & 2024-10-21 05:07:56 & Trigger & -- & 0.5--4.0 & $1.48_{-1.22}^{+1.24}$ & $\left(3.3_{-1.6}^{+4.8}\right)\times10^{-10}$ & \cite{Hu2024GCN} \\
        \emph{Swift} & XRT & 2024-10-24 20:58:09 & 3.7243 & 2.58 & 0.3--10 & 2.3 (fixed) & $<2.5\times10^{-13}$ & This work \\
        \emph{Swift} & XRT & 2024-10-25 15:30:26 & 4.47828 & 1.38 & 0.3--10 & 2.3 (fixed) & $<3.7\times10^{-13}$ & This work \\
        \emph{Swift} & XRT & 2024-10-29 09:32:45 & 8.34581 & 4.37 & 0.3--10 & 2.3 (fixed) & $\left(1.2_{-0.4}^{+0.5}\right)\times10^{-13}$ & This work \\
        \emph{Swift} & XRT & 2024-11-03 02:25:34 & 13.288 & 3.11 & 0.3--10 & 2.3 (fixed) & $<1.7\times10^{-13}$ & This work \\
        \emph{Chandra} & ACIS-S & 2024-11-04 02:51:52 & 13.97409 & 10.0 & 0.3--10 & $1.33\pm0.65$ & $\left(6.5^{+3.1}_{-2.0}\right)\times10^{-14}$ & This work \\ 
         &  & &  &  &  & 2.0 (fixed) & $\left(6.93_{-1.32}^{+1.52}\right)\times10^{-14}$ & This work \\ 
        \emph{Swift} & XRT & 2024-11-05 20:24:10 & 15.64207 & 1.0 & 0.3--10 & 2.3 (fixed) & $<4.9\times10^{-13}$ & This work \\
        \emph{Swift} & XRT & 2024-11-07 02:15:13 & 17.11778 & 3.2 & 0.3--10 & 2.3 (fixed) & $<1.8\times10^{-13}$ & This work \\
        \emph{Swift} & XRT & 2024-11-15 19:08:44 & 25.84677  & 1.7 & 0.3--10 & 2.3 (fixed) & $\left(7.2_{-3.7}^{+5.5}\right)\times10^{-14}$ & This work \\ 
        \emph{XMM-Newton} & MOS2 & 2025-01-18 13:03:43 & 89.34 & 39.9 & 0.3--10 & -- & $<2.83\times10^{-15}$ & \cite{Shu2025} \\ \hline
        \multicolumn{9}{c}{\emph{Einstein Probe satellite}} \\ \hline
        Einstein Probe & FXT & 2024-10-22 17:43:00 & 1.54  & 3.0 & 0.3--10 & 1.83$\pm$0.08 & $\left(2.10^{+0.32}_{-0.28}\right)\times10^{-13}$ & \cite{Shu2025} \\
        Einstein Probe & FXT & 2024-10-23 17:46:00 & 2.54  & 6.0 & 0.3--10 & -- & $\left(2.0^{+0.19}_{-0.18}\right)\times10^{-13}$ & \cite{Shu2025} \\
        Einstein Probe & FXT & 2024-10-24 19:25:00 & 3.61  & 6.0 & 0.3--10 & -- & $\left(1.45^{+0.18}_{-0.16}\right)\times10^{-13}$ & \cite{Shu2025} \\
        Einstein Probe & FXT & 2024-10-25 17:51:00 & 4.54  & 8.6 & 0.3--10 & -- & $\left(1.41^{+0.14}_{-0.15}\right)\times10^{-13}$ & \cite{Shu2025} \\
        Einstein Probe & FXT & 2024-10-27 14:47:00 & 6.41  & 6.1 & 0.3--10 & -- & $\left(1.66^{+0.20}_{-0.18}\right)\times10^{-13}$ & \cite{Shu2025} \\
        Einstein Probe & FXT & 2024-10-29 08:28:00 & 8.15  & 5.6 & 0.3--10 & -- & $\left(2.09^{+0.25}_{-0.18}\right)\times10^{-13}$ & \cite{Shu2025} \\
        Einstein Probe & FXT & 2024-11-02 15:04:00 & 12.43  & 2.1 & 0.3--10 & -- & $\left(0.91^{+0.26}_{-0.22}\right)\times10^{-13}$ & \cite{Shu2025} \\
        Einstein Probe & FXT & 2024-11-04 07:08:00 & 14.10  & 6.2 & 0.3--10 & -- & $\left(0.79^{+0.14}_{-0.13}\right)\times10^{-13}$ & \cite{Shu2025} \\
        Einstein Probe & FXT & 2024-11-07 12:04:00 & 17.30  & 4.8 & 0.3--10 & -- & $\left(0.87^{+0.18}_{-0.15}\right)\times10^{-13}$ & \cite{Shu2025} \\
        Einstein Probe & FXT & 2024-11-20 15:41:00 & 30.45  & 5.6 & 0.3--10 & -- & $\left(0.47^{+0.13}_{-0.11}\right)\times10^{-13}$ & \cite{Shu2025} \\
        Einstein Probe & FXT & 2024-11-30 19:13:00 & 40.60  & 8.9 & 0.3--10 & -- & $\left(0.31^{+0.10}_{-0.09}\right)\times10^{-13}$ & \cite{Shu2025} \\
        Einstein Probe & FXT & 2025-01-08 10:36:19 & 79.24  & 8.7 & 0.3--10 & -- & $<1.82\times10^{-14}$ & \cite{Shu2025} \\

    \hline
    \end{tabular}
    }
    \caption{X-ray detections of EP~241021a used in this work. \emph{Columns 1 and 2:} Telescope and instrument per observation, respectively. \emph{Column 3 and 4:} start date of the observation and days after the X-ray trigger, respectively. \emph{Columns 5 and 6:} Exposure time and energy band per observation, respectively. \emph{Column 7:} photon index computed or assumed. \emph{Column 8:} Unabsorbed flux. \emph{Column 9:} Reference.}
    \label{tab:x-rays}
\end{table*}

\begin{table*}
    \centering
    \resizebox{0.8\textwidth}{!}{
    \begin{tabular}{llcccc}
        \hline\hline
        Telescope & Start date (UT) & Since trigger (d) & Frequency (GHz) & Flux density (mJy) & Reference \\ 
        (1) & (2) & (3) & (4) & (5) & (6) \\ \hline
    ATCA & 2024-10-29 & 8.0 & 5.0 & 0.350$\pm$0.029 & \cite{Shu2025} \\
    ATCA & 2024-10-29 & 8.0 & 5.5 & 0.382$\pm$0.024 & \cite{Shu2025} \\
    ATCA & 2024-10-29 & 8.0 & 6.0 & 0.407$\pm$0.029 & \cite{Shu2025} \\
    ATCA & 2024-10-29 & 8.0 & 9.0 & 0.453$\pm$0.026 & \cite{Shu2025} \\
    ATCA & 2024-10-29 & 8.0 & 9.5 & 0.467$\pm$0.028 & \cite{Shu2025} \\ 
    
    ATCA & 2024-11-08 8:52:55 & 18.3 & 5.5 & 0.434$\pm$0.022 & \cite{Yadav2025} \\
    ATCA & 2024-11-08 8:52:55 & 18.3 & 9.0 & 0.543$\pm$0.024 & \cite{Yadav2025} \\
    ATCA & 2024-11-08 8:52:55 & 18.3 & 17.0 & 0.366$\pm$0.035 & \cite{Yadav2025} \\
    ATCA & 2024-11-11 & 21.1 & 2.1 & 0.256$\pm$0.042 & \cite{Gianfagna2025} \\ 
    e-MERLIN & 2024-11-09 & 19.5 & 5.0 & 0.614$\pm$0.164 & \cite{Gianfagna2025} \\ 
    
    ATCA & 2024-11-15 & 25.1 & 5.5 & 0.842$\pm$0.044 & \cite{Gianfagna2025} \\
    ATCA & 2024-11-15 & 25.1 & 9.0 & 0.653$\pm$0.034 & \cite{Gianfagna2025} \\
    ATCA & 2024-11-15 & 25.1 & 16.7 & 0.341$\pm$0.029 & \cite{Gianfagna2025} \\
    ATCA & 2024-11-15 7:36:25 & 25.3 & 17.0 & 0.347$\pm$0.041 & \cite{Yadav2025} \\ 
    ATCA & 2024-11-15 & 25.1 & 21.0 & 0.285$\pm$0.092 & \cite{Gianfagna2025} \\

    ATCA & 2024-11-21 & 31.6 & 2.1 & 0.327$\pm$0.033 & \cite{Gianfagna2025} \\
    ATCA & 2024-11-21 & 31.6 & 5.5 & 1.07$\pm$0.09 & \cite{Gianfagna2025} \\
    ATCA & 2024-11-21 & 31.6 & 9.0 & 1.1$\pm$0.07 & \cite{Gianfagna2025} \\ 

    VLBA & 2024-11-28 & 37.0 & 6.2 & 0.774$\pm$0.072 & \cite{Shu2025} \\
    VLBA & 2024-11-40 & 39.0 & 8.4 & 0.762$\pm$0.081 & \cite{Shu2025} \\
    ATCA & 2024-11-24 7:01:00 & 34.1 & 17.0 & 0.450$\pm$0.046 & \cite{Yadav2025} \\
    ATCA & 2024-11-24 & 34.1 & 16.7 & 0.393$\pm$0.061 & \cite{Gianfagna2025} \\
        \hline
    \end{tabular}
    }
    \caption{Radio observations of EP241021a used in this work. \emph{Column 1:} telescope per observation. \emph{Columns 2 and 3:} start day of the observation and the day since the X-ray trigger, respectively. \emph{Columns 4, 5 and 6:} frequency, flux density per observation (mJy units), and references, respectively.}
    \label{tab:radio}
\end{table*}

\begin{table*}
    \centering
    \resizebox{0.8\textwidth}{!}{
    \begin{tabular}{lllccc}
        \hline\hline
        Telescope & Instrument & Start date (UT) & Since trigger (d) & Exp. time (s) & Grism/Grating \\ 
        (1) & (2) & (3) & (4) & (5) & (6) \\ \hline
        VLT & FORS2 & 2024-10-24 05:19:51 & 3.02246 & 2$\times$1200 & GRIS\_300V \\
        GS & GMOS & 2024-10-25 02:40:34  & 3.9254 & 4$\times$900 &  R400  \\
        GS & GMOS & 2024-10-28 04:03:47  & 6.992 & 4$\times$1500 &  R400  \\
        GTC & OSIRIS+ & 2024-10-31 00:45:06 & 9.8505 & 3$\times$1800 & R500R \\
        GS & GMOS & 2024-10-31 06:27:26  & 10.085 & 4$\times$1200 &  R400  \\
        GTC & OSIRIS+ & 2024-11-09 23:16:22 & 19.7874 & 3$\times$1800 &  R300B \\
        VLT & MUSE & 2024-11-27 01:52:05 & 36.8956 & 4$\times$700 & -- \\
        \hline
    \end{tabular}
    }
    \caption{Spectroscopy obtained with various ground-based Telescopes for this Work.}
    \label{tab:spectroscopy}
\end{table*}

\begin{table}
    \centering
    \begin{tabular}{ccc}
    \hline\hline
    Epoch & Lines & $z$ \\ 
    (1) & (2) & (3) \\ \hline
    GMOS 1 & 3727, H$_\beta$, 4959, 5007\AA & 0.7486\\
    GMOS 2 & 3727, H$_\beta$, 4959, 5007\AA & 0.7486\\
    GMOS 3 & H$_\beta$, 5007\AA & 0.7487\\
    OSIRIS 1 & 3727, 4959, 5007\AA & 0.7488\\
    OSIRIS 2 & 3727, 4959, 5007\AA & 0.7481\\
    MUSE & H$_\beta$, 4959, 5007\AA & 0.7484\\
    \hline
    \end{tabular}
    \caption{Redshifts obtained from fitting the brightest emission lines per spectrum. \emph{Columns 1 and 2:} Spectroscopic epoch and emission lines used for measuring redshifts. \emph{Column 3:} Redshift obtained for each spectroscopic epoch. The mean redshift value from fitting emission lines is $\overline{z}=$0.7485}
    \label{tab:redshifts}
\end{table}

\begin{table*}
    \centering
    \begin{tabular}{lllllllll}
    \hline\hline
    Epoch (days) & $\log(\nu_m/{\rm Hz})$ & $\log(\nu_c/{\rm Hz})$ & $F_{\rm max}$ (mJy) & $p$ & $T_{\rm BB}$ ($\times10^3$~K) & $R_{\rm BB}$ ($\times10^{15}$~cm) & BIC \\ 
    (1) & (2) & (3) & (4) & (5) & (6) & (7) & (8) \\ \hline
    2.8 & $12.34_{-0.37}^{+0.16}$ & $18.21_{-1.49}^{+0.79}$ & $2.18_{-1.12}^{+0.97}$ & $2.96_{-0.18}^{+0.20}$ & -- & -- & $-70.59$ \\  
    8.3 & $12.5_{-0.03}^{+0.0}$ & $18.14_{-1.26}^{+0.86}$ & $2.6\pm0.17$ & $3.01\pm0.03$ & -- & -- & $294.85$\\
    18.2 & $10.22_{-0.09}^{+0.11}$ & $17.94_{-1.76}^{+1.05}$ & $0.54\pm0.05$ & $2.33_{-0.16}^{+0.12}$ & $8.93_{-0.82}^{+0.68}$ & $5.29_{-0.85}^{+0.92}$ &  $-284.82$ \\    
    24.5 & $9.75_{-0.25}^{+0.27}$ & $18.0_{-1.97}^{+0.99}$ & $1.14_{-0.38}^{+0.62}$ & $2.3_{-0.22}^{+0.13}$ & $5.6_{-1.60}^{+4.32}$ & $10.81_{-10.79}^{+15.6}$ &  $-312.29$ \\
    31.8 & $10.32_{-0.14}^{+0.18}$ & $18.09_{-2.06}^{+0.91}$ & $1.01_{-0.15}^{+0.20}$ & $2.5_{-0.21}^{+0.49}$ & $8.52_{-1.37}^{+1.26}$ & $4.56_{-4.35}^{+1.92}$ &  $-193.44$\\
    37.5 & $9.93_{-0.43}^{+0.32}$ & $17.14_{-0.63}^{+0.86}$ & $0.99_{-0.31}^{+1.18}$ & $2.38_{-0.14}^{+0.25}$ & $5.28_{-1.85}^{+3.86}$ & $12.81_{-9.86}^{+38.82}$ & $-201.03$ \\
    \hline
    \end{tabular}
    \caption{Posteriors of the models obtained fitting the SED of the transient at different epochs (see \S\ref{sec:fen_SED}). The parameters are at 99\% confidence level.}
    \label{tab:posteriors}
\end{table*}



\bsp	
\label{lastpage}
\end{document}